\documentclass[review, preprint,12pt]{elsarticle} %
\usepackage{amssymb}
\usepackage{amsmath,amsfonts,amssymb,bm,accents}
\usepackage{hyperref}
\usepackage{mathrsfs}
\usepackage{graphicx}
\usepackage{epstopdf}
\usepackage{float}
\usepackage{caption}
\usepackage{subcaption}
\usepackage{bbm}
\usepackage{mathrsfs}
\usepackage{cleveref}
\usepackage{soul}
\usepackage{accents}
\usepackage{graphicx}
\usepackage[export]{adjustbox}
\usepackage{xcolor}
\usepackage{courier} % To write in Courier font (to write code basically)
\usepackage{listings} % To include code
\usepackage{tabu} % To increase the vertical space between cells in tables
\usepackage{booktabs}
\usepackage{longtable}
\usepackage{changepage} % To use the adjustwidth feature that allows pushin wide tables to the left margin
\usepackage[margin=2cm]{geometry}% TO PLAY WITH THE MARGINS
\biboptions{sort&compress} %To make [1,2,3] be [1-3]
\usepackage[section]{placeins}
\usepackage[font={small}]{caption}

\usepackage{siunitx}
\usepackage[export]{adjustbox}

\newcommand{\beq}{\begin{equation}}
\newcommand{\eeq}{\end{equation}}
\newcommand{\bea}{\begin{eqnarray}}
\newcommand{\eea}{\end{eqnarray}}

\newcommand{\dd}{\,{\rm d}}

\usepackage{lineno}

\journal{Materials and Structures}

\makeatletter
\def\@author#1{\g@addto@macro\elsauthors{\normalsize%
    \def\baselinestretch{1}%
    \upshape\authorsep#1\unskip\textsuperscript{%
      \ifx\@fnmark\@empty\else\unskip\sep\@fnmark\let\sep=,\fi
      \ifx\@corref\@empty\else\unskip\sep\@corref\let\sep=,\fi
      }%
    \def\authorsep{\unskip,\space}%
    \global\let\@fnmark\@empty
    \global\let\@corref\@empty  %% Added
    \global\let\sep\@empty}%
    \@eadauthor={#1}
}
\makeatother

\DeclareUnicodeCharacter{2212}{-}

\begin{document}

\begin{frontmatter}



\title{Predicting the impact of water transport on carbonation-induced corrosion in variably saturated reinforced concrete}


\author[IC]{Ev\v{z}en Korec}
%\ead{e.korec20@imperial.ac.uk}

\author[UP]{Lorenzo Mingazzi}

\author[UP]{Francesco Freddi}

\author[Oxf,IC]{Emilio Mart\'{\i}nez-Pa\~neda\corref{cor1}}
\ead{emilio.martinez-paneda@eng.ox.ac.uk}

\address[IC]{Department of Civil and Environmental Engineering, Imperial College London, London SW7 2AZ, UK}

\address[UP]{Department of Engineering and Architecture, Università degli Studi di Parma, Parco Area delle Scienze 181/A, Parma 43124, Italy}

\address[Oxf]{Department of Engineering Science, University of Oxford, Oxford OX1 3PJ, UK}

\cortext[cor1]{Corresponding author.}

\begin{abstract}
A modelling framework for predicting carbonation-induced corrosion in reinforced concrete is presented. The framework constituents include a new model for water transport in cracked concrete, a link between corrosion current density and water saturation, and a theory for characterising concrete carbonation. The theoretical framework is numerically implemented using the finite element method and model predictions are extensively benchmarked against experimental data. The results show that the model is capable of accurately predicting carbonation progress, as well as wetting and drying of cracked and uncracked concrete, revealing a very good agreement with independent experiments from a set of consistent parameters. In addition, insight is gained into the evolution of carbonation penetration and corrosion current density under periodic wetting and drying conditions. Among others, we find that cyclic wetting periods significantly speed up the carbonation progress and that the induced corrosion current density is very sensitive to concrete saturation.\\    
\end{abstract}


\end{frontmatter}

%\linenumbers
%\begin{linenumbers}

\section{Introduction}
\label{Introduction}
This study presents a new model for predicting carbonation corrosion in reinforced concrete subjected to variable moisture saturation. Carbonation results from the penetration of atmospheric carbon dioxide (CO$_{\text{2}}$) into concrete, leading to a series of chemical reactions involving the transformation of calcium hydroxide into calcium carbonate, which causes acidification of the basic concrete pore solution \citep{Poursaee2016a, Angst2018a, Andrade2019}. This pH drop is known to destabilise the protective passive layer on the surface of steel rebars, triggering rebar corrosion \citep{Poursaee2016a}. Corrosion of steel rebars is the main deterioration mechanism of concrete structures, being responsible for the premature degradation of 70-90\% of them \cite{Gehlen2011-za, British_Cement_Association_BCA1997-jj}. Also, carbonation can negatively affect the freeze-thaw resistance and can lead to the shrinkage of concrete \citep{Angst2020a}. On the other hand, it can also have some positive effects such as the enhancement of mechanical properties of concrete made from Portland cement, crack self-healing \cite{Yang2022}, increased resistance to low-temperature sulfate attack \cite{Zhang2021a} or the removal of carbon dioxide from the atmosphere \citep{Angst2020a}. \\ 

Concrete carbonation has been one of the major concerns for researchers aiming at reducing carbon emissions by developing new low-carbon 
binders, where a significant portion of clinker (ordinary Portland cement) is replaced with supplementary cementitious materials such as fly-ash, limestone, and geopolymers. Although the resulting material has a lower environmental footprint, lower calcium hydroxide content leads to a reduced concrete pH buffer capacity and a higher susceptibility to carbonation \citep{Angst2020a, Dhandapani2022, VonGreve-Dierfeld2020, Leemann2018}.\\ 

Concrete water saturation and microstructure play a dominant role in the carbonation process \citep{Leemann2015a,Angst2020a,Poursaee2016a}. Low levels of water saturation facilitate the penetration of atmospheric carbon dioxide into concrete but the reaction proceeds in pore solution and thus its rate increases with concrete water saturation \citep{Poursaee2016a}. Also, carbonation affects the concrete microstructure and consequently the water permeability of concrete in a complex, and so far not entirely understood, way. In Portland cement, carbonation-induced pore clogging was found to prevail over micro-cracking, reducing permeability, while for concretes made from blended cements the opposite was observed \citep{Auroy2015}.

For these reasons, the interaction of carbonation and water transport has been intensively studied and a number of models have been proposed. Although some early models initially did not explicitly consider water transport \citep{Papadakis1991}, many coupled models have been introduced since \cite{Saetta1993a,Steffens2002,Isgor2004,Song2006,Bary2004,OmikrineMetalssi2020, Hwang2020}. \citet{Seigneur2020,Seigneur2022} conducted detailed reactive transport modelling of carbonation considering the role of various mineral phases of cement paste. \citet{Bretti2022} numerically investigated the interplay between carbonation in Portland cement and pore water content, as well as the porosity variation resulting from the carbonation process. \citet{Nguyen2015} formulated a coupled model that explicitly resolved the mesoscale nature of concrete. Although diffusion is typically the dominant transport mechanism of carbon dioxide in concrete \cite{Forsdyke2023}, advection under high hydrostatic pressure could also play a role, for example in the case of radioactive waste disposal underground concrete structures, as studied by \citet{Phung2016}. \citet{Kari2014} used a coupled model to show that extrapolation based on a linear diffusion equation was too simplistic to describe the kinetics involved in long-term carbonation. The combined effect of carbonation and chloride attack was investigated by \citet{Zhu2016a,Shen2019,Meijers2005,Li2018} and \citet{Xie2021}. Recently, \citet{Freddi2022} coupled carbonation and fracture by means of a phase field model, allowing for the investigation of the interaction between carbonation and the corrosion-induced cracking process that is triggered by carbonation, but without incorporating the role of moisture. Developing models capable of resolving the interplay between cracks, carbonation, water content and corrosion is critical to deliver service life predictions; yet this is an area that remains to be explored. In particular, there is a need for models capable of capturing the role that cracks play in accelerating water transport, as this is arguably a dominant contribution \cite{Grassl2009}.\\ 

In this work, we present a new theoretical and computational framework that accounts for: (i) the transport of water through bulk and cracked concrete, (ii) the interplay between water saturation and corrosion current density, and (iii) concrete carbonation. The proposed coupled water transport and carbonation reactive transport model is combined with a phase field description of cracks and numerically implemented using the finite element method. The proposed theory and details of the finite element implementation are given in Section \ref{Sec:Theory}. Then, in Section \ref{Sec:Results}, model predictions are benchmarked against experimental measurements of relative water mass loss, water saturation ratio, water penetration contours and carbonation penetration depths, for both drying and wetting of cracked and uncracked concrete. In all cases, a very good agreement between simulations and test data is observed. Finally, insight is gained into the role of variable moisture saturation and cracks on the corrosion current density. The manuscript ends with concluding remarks in Section \ref{Sec:Conclusions}. 

\section{Theory and computational implementation}
\label{Sec:Theory}

In this section, the underlying theory of the proposed model is presented, together with brief details of the numerical implementation. Firstly, the water transport model resolving water saturation in the concrete pore space is presented in Section \ref{subSec:WatTransMod}. This is followed by the description of the interplay between corrosion current density and water saturation, following the work of \citet{Stefanoni2019} (Section \ref{subSec:CorrCurrMod}). Then, in Section \ref{subSec:CarbMod}, the concrete carbonation model is provided. Finally, an overview of the governing equations of the model is given in Section \ref{Sec:govEq}.\\ 

\noindent \emph{Notation}. Scalar quantities are denoted by light-faced italic letters, e.g. $\phi$, Cartesian vectors by upright bold letters, e.g. $\mathbf{u}$, and Cartesian second- and higher-order tensors by bold italic letters, e.g. $\bm{K}$. The symbol $ \bm{1} $ represents the second-order identity tensor. Finally, $ \bm{\nabla} $ and $ \bm{\nabla} \cdot $ respectively denote the gradient and divergence operators.

\subsection{A model for characterising water transport in cracked concrete}
\label{subSec:WatTransMod}

Moisture in concrete can be transported in both liquid and gas form (i.e., as vapour). The analysis of \citet{mainguy2001role} revealed that during the drying of weakly permeable materials such as concrete, the liquid transport mechanism dominates over the contribution of gas transport. In addition, our numerical simulations showed that water vapour transport is also negligible during wetting by liquid water. Accordingly, we proceed to neglect the effects of water vapour transport, evaporation and gas pressure. With these assumptions, following \citet{mainguy2001role}, the transport equation of moisture in the concrete domain $\Omega^{c} $ can be formulated as      
\begin{equation}\label{transEq1}
\frac{\partial}{\partial t}\left(\theta \rho_l S_l \right)=-\bm{\nabla}\cdot\left(\theta S_l \rho_l \mathbf{v}_l\right)
\end{equation}
where $t$ denotes time, $ \theta $ is the porosity of concrete, which changes in time with carbonation progress, $ \rho_l $ denotes the density of water, $ S_{l} $ is the unknown liquid saturation ratio, and $ \mathbf{v}_l $ is the velocity of the liquid phase. Based on Darcy's law \cite{mainguy2001role}, the volume flux can be related to the gradient of liquid pressure as 
\begin{equation}\label{transEq2}
\theta S_l \mathbf{v}_l = - \frac{k_{r}}{\eta}\bm{K}\cdot\nabla p_l 
%= - \frac{k_{r}}{\eta}\dfrac{\mathrm{d}p_l(S_l)}{\mathrm{d}S_l}(\bm{K}\cdot \nabla S_l(t)
\end{equation}
Here, $ k_{r} $ is the relative permeability, $\eta $ is the dynamic viscosity of liquid water, $\bm{K}$ is the intrinsic permeability tensor, and $p_l$ is the liquid pressure. The difference between gas pressure $p_g$ and liquid pressure $p_l$ is the capillary pressure $ p_c = p_g - p_l$. Since gas pressure is usually negligible, one can assume that $ p_l = -p_c $ and, accordingly, Eq. (\ref{transEq2}) can be expressed in terms of the capillary pressure as 
\begin{equation}\label{transEq3}
\theta S_l \mathbf{v}_l = \frac{k_{r}}{\eta}\bm{K}\cdot\nabla p_c 
%= \frac{k_{r}}{\eta}\dfrac{\mathrm{d}p_c(S_l)}{\mathrm{d}S_l}\bm{K}(t)\cdot \nabla S_l
\end{equation}
The balance equation (\ref{transEq1}) is written in terms of the liquid saturation ratio $S_l$ as the primary unknown, but the transport law (\ref{transEq3}) uses the gradient of capillary pressure as the driving force of the transport process. 
For a given porous material, the capillary pressure $p_c$ can be expressed as a function of the liquid saturation ratio $S_l$. This function $p_c(S_l)$ is known as the capillary curve and depends on the material. The experimentally determined capillary curve is commonly fitted with the expression originally proposed by \citet{van1980closed}, which reads
\begin{equation}\label{VanGenuchten}
p_c\left(S_l\right)=\alpha\left(S_l^{-\beta}-1\right)^{1-1 / \beta}
\end{equation}    
where $\alpha$ and $\beta$ are material parameters. It should be noted that if the material undergoes periodic wetting and re-drying, the capillary curve changes because of sorption hysteresis effects. Currently, such effects are not considered in this model, but they could be readily included, as done for example by \citet{Zhang2015a}. 
\\
\\
Substituting (\ref{transEq3}) into (\ref{transEq1}), and assuming incompressibility of water (i.e., constant $\rho_l$), we obtain
\begin{equation}\label{transEq4}
\frac{\partial}{\partial t}\left(\theta S_l\right)=-\bm{\nabla}\cdot\left(\frac{k_{r}}{\eta}\dfrac{\mathrm{d}p_c}{\mathrm{d}S_l}\bm{K}\cdot \nabla S_l\right) \,\,\,\,\,\, \text{ in } \Omega^{c}  
\end{equation}     
\noindent The capillary pressure is also related to the relative humidity $ h_r $ by the Kelvin law, which reads
\begin{equation}\label{KelvinLaw}
p_c = -\rho_l \frac{R T}{M_l} \ln h_r
\end{equation}
where $R$ is the gas constant, $T$ is the absolute temperature and $M_l$ is the molar mass of water. The Kelvin law (\ref{KelvinLaw}) can be used in combination with the capillary curve (\ref{VanGenuchten}) to calculate the initial condition for $S_l$ and the boundary condition for $S_l$ on surfaces where the relative humidity is known. 
\\

By substituting equation (\ref{KelvinLaw}) into (\ref{VanGenuchten}), we obtain a relation between relative humidity and saturation ratio that is commonly referred to as a sorption isotherm,
\begin{equation}\label{Isotherm}
S_l(h_r) = \left(1 + \left(-\rho_l \frac{RT}{M_l \alpha} \ln h_r \right)^{\beta/(\beta-1)}\right)^{-1/\beta}
\end{equation}
The sorption isotherm is calibrated from experimental data by an appropriate choice of parameters $\alpha$ and $\beta$. It remains to formulate the expressions for the relative permeability $ k_r $ and the intrinsic permeability $ \bm{K} $. The relative permeability $ k_r $ determines the ratio of the effective permeability of the liquid water to the total permeability of the porous material. Based on Mualem’s model \cite{mualem1976new}, which predicts hydraulic conductivity from the statistical pore-size distribution, and employing the experimentally-fitted capillary curve (\ref{VanGenuchten}), \citet{van1980closed} derived the following expression for relative permeability $ k_r $:
\begin{equation}\label{relPerm}
k_r = \sqrt{S_{l}}\left[1-\left(1-S_l^{\beta}\right)^{1/\beta}\right]^2
\end{equation}       
\\
We proceed now to incorporate the role of cracks. Intrinsic permeability is much higher in cracks than in an undamaged material. Inspired by the phase field hydraulic fracture literature \cite{Miehe2015b,Wilson2016,Heider2020}, we express the intrinsic permeability tensor $\bm{K}$ as the sum of the contributions of isotropic bulk permeability $ \bm{K}_{m} $ and an anisotropic cracked permeability $\bm{K}_{c}$:
\begin{equation}\label{perm1}
\bm{K}=\bm{K}_{m}+\bm{K}_{c} 
\end{equation}
The isotropic bulk permeability $ \bm{K}_{m} $ is here estimated from porosity, following \citet{Zhang2022};
\begin{equation}\label{isoAbsPerm}
\bm{K}_{m} = k \bm{1} = \frac{\theta^3}{C \tau^2 \rho_s^2} \bm{1}
\end{equation}
where $\textbf{1}$ is the second-order identity tensor, $ \rho_s $ is the density of the dry material and $ C $ is a fitting constant. Because tortuosity can be estimated with the Bruggeman relation \cite{Zhang2022} as $\tau=\theta^{-2.5}$, Eq. (\ref{isoAbsPerm}) can be reformulated as  
\begin{equation}\label{isoAbsPerm1}
\bm{K}_{m} = k \bm{1} = \frac{\theta^8}{C \rho_s^2} \bm{1}
\end{equation}
Assuming laminar flow in the crack of opening $ w $, we define the anisotropic cracked permeability $\bm{K}_{c}$ as:      
\begin{equation}\label{perm2}
\bm{K}_c= \phi \frac{w^2}{12}\left(\bm{1}-\textbf{n}_{\phi} \otimes \textbf{n}_{\phi}\right) 
\end{equation}
where $\phi$ is a phase field variable characterizing concrete damage \cite{Wu2021,Kristensen2021, Miehe2010, Amor2009, Bourdin2000}. Akin to a damage variable, the phase field order parameter changes from $\phi = 0$ in uncracked concrete to $\phi = 1$ in fully cracked material points. The core idea of the phase-field approach is to replace the sharp crack geometry representing a discontinuity in the damage field with the phase-field variable $\phi$ varying sufficiently smoothly over the crack process zone. In mathematical terms, this is achieved by adding an additional term to the total potential energy density functional which contains $|\nabla \phi|^2$; the reader is referred to (e.g.) Ref. \cite{Korec2023, Carrara2020} for derivation details. This ensures that large gradients of $\phi$ are penalised and $\phi$ is maximal in the centre of the process zone and sufficiently smoothly decreases towards its boundaries. Phase field fracture models have gained increasing attention in recent years and their success has been extended to the modelling of concrete cracking 
\cite{Navidtehrani2022,Huang2022,Wu2017,Wu2018a}, including in combination with reactive transport modelling \cite{Korec2023,Pundir2023,Freddi2022,Wu2022,korec2024phase}. Here, our purpose is to exploit the phase field regularisation of cracks to define regions of initial damage and describe the special characteristics of transport and carbonation within them. Hence, the phase field distribution is given by,
\begin{equation}\label{pfEq}
\phi+ \ell^2 \bm{\nabla}\cdot \nabla \phi = 0  
\end{equation}
Where $\ell$ is a phase field length scale that controls the width of the fracture process zone. To achieve mesh-insensitive results, the element size $H_{e}$ in the process zone must be 5–7 times smaller than $\ell$ \cite{Kristensen2021}. It remains to define the opening of the crack $ w $. To this end, we define $\textbf{n}_{\phi}=\nabla \phi /|\nabla \phi|$ as a normalised vector pointing perpendicularly to the crack, such that the operator $\bm{1}-\textbf{n}_{\phi} \otimes \textbf{n}_{\phi}$ projects the enhancement of permeability only in the direction of the crack. Then, the crack opening $ w $ is approximated as
\begin{equation}\label{crOpen}
w=\left\{\begin{array}{l}
0\text{ if } \phi<\phi_t \\
w_{cr} \text{ if } \phi \geq \phi_t
\end{array}\right.
\end{equation}
\noindent where $w_{cr}$ is the crack width and $ \phi_t $ is the threshold defining the crack contours as only the part of the regularised phase-field profile represents an opened crack. In the further described case studies, the position of cracks and their widths were known and $w_{cr}$ were thus prescribed directly. However, if cracks result from a coupled mechanical problem, $w_{cr}$ is not known a priori. In such cases, it can be calculated as $w_{cr} = H_e \left(\mathbf{n}_{\phi} \cdot \bm{\varepsilon} \cdot \mathbf{n}_{\phi} \right)$  where $\bm{\varepsilon} = \nabla_s \textbf{u}$ is the small strain deformation tensor, with $\textbf{u}$ being the displacement vector \cite{Wilson2016,Miehe2015b}.

\subsection{Linking the corrosion current density to the water saturation}
\label{subSec:CorrCurrMod}

The water saturation ratio $S_l$, as determined by Eq. (\ref{transEq4}), can be related to the corrosion current density, given the dependency of the latter on pore structure and moisture state \cite{Stefanoni2019}. The pore structure has a twofold effect. Firstly, it influences the transport of released ferrous ions into the porosity because the pore structure can act as a diffusion constraint, limiting the corrosion current density. Secondly, together with the moisture state, the pore structure determines the area of the steel surface that is in contact with moisture, which is a necessary condition for the corrosion process to proceed \cite{Angst2017,Wong2022}. Based on these ideas, we follow \citet{Stefanoni2019} and define the corrosion current density $i_{c}$ as a function of the water saturation ratio $S_l$ and the material porosity $\theta$, such that
\begin{equation}\label{corrDenAngst}
i_{c}=i_{max} \frac{1}{2}\left(1+\frac{\left(\theta-\theta_{crit}\right)}{\sqrt{k+\left(\theta-\theta_{crit}\right)^2}}\right) S_{l}
\end{equation}
Here, $ k $ is a non-dimensional fitting parameter and $\theta_{crit}$ is a critical porosity level, with both influencing the shape of the function $ i_{c}(\theta) $. Specifically, $\theta_{crit}$ specifies the inflection point of $ i_{c}(\theta) $. The maximum corrosion current density, $ i_{max} $, which depends on the composition of the pore solution, acts as an asymptotic value of (\ref{corrDenAngst}) and represents the maximum corrosion rate in a completely open system, i.e. solution, without any diffusion constraint by the pore structure. Following the experimental calibration by \citet{Stefanoni2019}, we adopt $ i_{max} = 3.7$ \unit{\micro\ampere\per\centi\metre^2}, $k = 10^{-3}$ and $\theta_{crit} = 0.185$.

\subsection{A model for concrete carbonation}
\label{subSec:CarbMod}

The carbonation of Portland cement concrete involves several stages \cite{Papadakis1991,Isgor2004}. Firstly, gaseous (g) carbon dioxide present in the air gradually penetrates concrete by means of diffusion through the concrete pore space. The gaseous carbon dioxide then dissolves in concrete pore solution (aq) through the following reactions 
\begin{equation}\label{reacCarbAcid}
\begin{aligned}
& \mathrm{H}_2 \mathrm{O}+\mathrm{CO}_2(\mathrm{g}) \rightarrow \mathrm{HCO}_3^{-}(\mathrm{aq})+\mathrm{H}^{+}(\mathrm{aq}) \\
& \mathrm{HCO}_3^{-}(\mathrm{aq}) \rightarrow \mathrm{CO}_3^{2-}(\mathrm{aq})+\mathrm{H}^{+}(\mathrm{aq})
\end{aligned}
\end{equation}

The concrete pore solution also contains calcium cations ($\mathrm{Ca}^{2+}$) emerging from the following dissolution reaction of solid (s) calcium hydroxide ($\mathrm{Ca}(\mathrm{OH})_2$) from hardened cement paste    
\begin{equation}\label{reacCemPasDiss}
\mathrm{Ca}(\mathrm{OH})_2(\mathrm{s}) \rightarrow \mathrm{Ca}^{2+}(\mathrm{aq})+2 \mathrm{OH}^{-}(\mathrm{aq})
\end{equation}
Then, calcium cations ($\mathrm{Ca}^{2+}$) and carbonate anions ($\mathrm{CO}_3^{2-}$) undergo the following neutralization reaction 
\begin{equation}\label{reacNeutr}
\mathrm{Ca}^{2+}(\mathrm{aq})+2 \mathrm{OH}^{-}(\mathrm{aq})+2 \mathrm{H}^{+}(\mathrm{aq})+\mathrm{CO}_3^{2-}(\mathrm{aq}) \rightarrow \mathrm{CaCO}_3(\mathrm{s})+2 \mathrm{H}_2 \mathrm{O}
\end{equation}
 and form calcium carbonate ($\mathrm{CaCO}_3$).
 
Accordingly, the carbonation process of Portland cement has two critical consequences. Firstly, emerging calcium carbonate gradually fills concrete porosity and secondly, calcium hydroxide ($\mathrm{Ca}(\mathrm{OH})_2$) is consumed by the carbonation reaction. Because calcium hydroxide is responsible for the basic character of the concrete pore solution (pH 12-13), the carbonation reaction has an acidifying effect, reducing the pH below 9. The acidification of the pore solution prevents the stable preservation of the protective passive layer on the surface, leading to the onset of corrosion \citep{Poursaee2016a}. It should be noted that the described series of reactions is valid only for concretes from Portland cement and may differ for blended cement such as blast furnace slag or fly ash cements.\\

To capture the carbonation process, we enrich our theoretical framework with a carbonation model based on the works by \citet{Freddi2022} and \citet{Isgor2004}. To this end, we describe the diffusion of carbon dioxide through a concrete domain $\Omega^{c} $ and the consumption of calcium hydroxide by
\begin{equation} \label{eq:CO2_diffusion}
\frac{\partial}{\partial t} \left(\theta \left(1 - S_l \right) c_{\mathrm{CO}_2} \right) - \bm{\nabla} \cdot \left( D_{\mathrm{CO}_2} \nabla c_{\mathrm{CO}_2} \right) = - \theta S_l R_n \,\,\,\,\,\, \text{ in } \Omega^{c} 
\end{equation} 
\begin{equation} \label{eq:CaOH_final_equation}
\frac{\partial}{\partial t} c_{\mathrm{Ca(OH)}_2} = -\theta S_l R_{n} \,\,\,\,\,\, \text{ in } \Omega^{c}   
\end{equation}  
where $c_{\mathrm{CO}_2}$ and $c_{\mathrm{Ca(OH)}_2}$ are the concentrations of carbon dioxide and calcium hydroxide, respectively. In addition, $D_{\mathrm{CO}_2}$ is the diffusivity of carbon dioxide in concrete and $R_{n}$ is the rate of neutralization reaction (\ref{reacNeutr}). The former is calculated as a function of the concrete porosity $\theta$ and the water saturation $S_l$ as,
\begin{equation} \label{eq:Diff_coeff_CO2} 
D_{\mathrm{CO}_2}= 1.64 \cdot 10^{-6} \left(\theta + (1 - \theta) \phi^{10} \right)^{1.8} (1 - S_l)^{2.2} 
\end{equation}

Diffusivity increases with porosity, but the higher the water saturation of pore space, the slower gaseous carbon dioxide can penetrate through concrete. The dependency on the phase field variable $\phi$ enriches the model to enable capturing the impact of enhanced diffusivity inside of cracks. For uncracked concrete ($\phi=0$) with 50$\%$ saturated porosity ranging between 10-20$\%$ of concrete volume, Eq. (\ref{eq:Diff_coeff_CO2}) predicts CO$_2$ diffusivity levels in the order of magnitude of $10^{-8}$ m$^2$s$^{-1}$, which agrees with the values measured by \citet{Papadakis1991}. The rate of the neutralization reaction (\ref{reacNeutr}) is expressed as 
\begin{equation} \label{eq:CO2_reaction}
R_n = HRT k_n c_{\mathrm{OH}^-}^{\mathrm{eq}} c_{\mathrm{CO}_2} c_{\mathrm{Ca(OH)}_2}
\end{equation}
where $H$ is the Henry constant for the dissolution of $\mathrm{CO}_2$ in water, $k_n$ is the reaction rate constant and $c_{\mathrm{OH}^-}^{\mathrm{eq}}$ is the $\mathrm{OH}^{-}$ equilibrium concentration. 

The progress of the carbonation process is tracked by carbonation front variable $\varphi \in \left[ 0,1 \right]$, which is calculated as  
\begin{equation} \label{eq:carbonation_front}
\varphi = 1 - \frac{c_{\mathrm{Ca(OH)}_2}}{c_{\mathrm{Ca(OH)}_2}^0}
\end{equation} 
with $c_{\mathrm{Ca(OH)}_2}^0$ being the initial calcium hydroxide concentration. A value of $\varphi = 0$ denotes uncarbonated concrete, while $\varphi = 1$ characterises fully carbonated concrete. The carbonation process affects the porosity of concrete and the pH of the pore solution. These are calculated based on the value of the carbonation front variable $\varphi$ as 
\begin{equation} \label{eq:material_variation}
\theta = \theta_0 + \varphi(\theta_{c} - \theta_0);
\end{equation}
\begin{equation} \label{eq:ph_eval}
\mathrm{pH} = 14 + \text{log}(2 \cdot 10^3 c_{\mathrm{Ca(OH)}_2}) 
\end{equation} 
In Eq. (\ref{eq:material_variation}), $ \theta_0 $ is the initial porosity of uncarbonated concrete and $\theta_{c} < \theta_0 $ is the porosity of fully carbonated concrete. Thus, changes in concrete porosity and pH are naturally captured by predicting the evolution of the concentrations of $\mathrm{CO}_2$ and $\mathrm{Ca(OH)}_2$, the primary variables of the carbonation model.

\subsection{Overview of the governing equations and details of the numerical implementation}
\label{Sec:govEq}
The system of governing equations for the coupled water transport and carbonation problem is given by
\begin{subequations}\label{summGovEq}
\begin{align} 
\frac{\partial}{\partial t}\left(\theta S_l\right) + \bm{\nabla}\cdot\left(\frac{k_{r}}{\eta}\dfrac{\mathrm{d}p_c}{\mathrm{d}S_l}\bm{K}\cdot \nabla S_l\right) &= 0 \,\,\,\,\,\, \text{ in } \Omega^{c} \label{eq:watTrans}\\
\frac{\partial}{\partial t} \left(\theta \left(1 - S_l \right) c_{\mathrm{CO}_2} \right) - \bm{\nabla} \cdot \left( D_{\mathrm{CO}_2} \nabla c_{\mathrm{CO}_2} \right) &= - \theta S_l R_n \,\,\,\,\,\, \text{ in } \Omega^{c}\label{eq:cdDiff}\\
\frac{\partial}{\partial t} c_{\mathrm{Ca(OH)}_2} &= -\theta S_l R_{n} \,\,\,\,\,\, \text{ in } \Omega^{c} \label{eq:chEvol}
\end{align}
\end{subequations}
With the primary unknowns being the water saturation ratio in the pore space $S_l$, the concentration of carbon dioxide $c_{\mathrm{CO}_2}$ and the concentration of calcium hydroxide $c_{\mathrm{Ca(OH)}_2}$. Dirichlet boundary conditions are prescribed on the boundaries of concrete exposed to atmospheric conditions (air or water); i.e. $S_l = \overline{s}$ and $c_{\mathrm{CO}_2} = \overline{c}$, while on protected boundaries zero flux of water or carbon dioxide is considered (i.e. $\textbf{n}\cdot\bm{K}\cdot \nabla S_l  = 0$ and $\textbf{n}\cdot D_{\mathrm{CO}_2} \nabla c_{\mathrm{CO}_2}  = 0$). Eq. (\ref{eq:chEvol}) does not contain any space derivatives and thus does not require any boundary conditions. The numerical implementation is carried out using the finite element method and the system of differential equations (\ref{eq:watTrans})-(\ref{eq:chEvol}) is solved using a fully implicit solution scheme. The concrete domain $\Omega^{c} $ is discretised with linear quadrilateral elements. All numerical simulations are performed using the open-source computing platform DEAL.II \citep{Arndt2021, Bangerth2007}. 

\section{Results}
\label{Sec:Results}

We proceed to showcase the ability of the model to predict experiments and provide new insight. To ensure consistency, a set of material parameters are first defined in Section \ref{Sec:modelParam}, which are subsequently used to predict the outcome of independent experiments with different boundary conditions. Specifically, the ability to replicate water transport during wetting and drying in uncracked concrete is validated with the experiments by \citet{Baroghel-Bouny1999} and \citet{Zhang2022} in Sections \ref{Sec:ResValWatTranDry} and \ref{Sec:ResValWatTranWet}, respectively. The ability of the model to capture the impact that cracks have on water transport is evaluated in Section \ref{Sec:ResValWatTranCrack}, by benchmarking model predictions against the experimental data of \citet{Michel2018}. In Section \ref{Sec:ResValCarb}, the coupling between water transport and carbonation is assessed by comparing predictions of carbonation depth under variable humidity with the testing data of \citet{Liu2020}. Finally, in Section \ref{Sec:ResCarbCorrDenWetDry}, the validated model is used to gain new insight into the evolution of carbonation and induced corrosion current density in cracked and uncracked samples subjected to cyclic wetting and drying. 

\subsection{Choice of model parameters}
\label{Sec:modelParam}

\begin{figure}[H]
	\centering
	\includegraphics[width=0.6\textwidth]{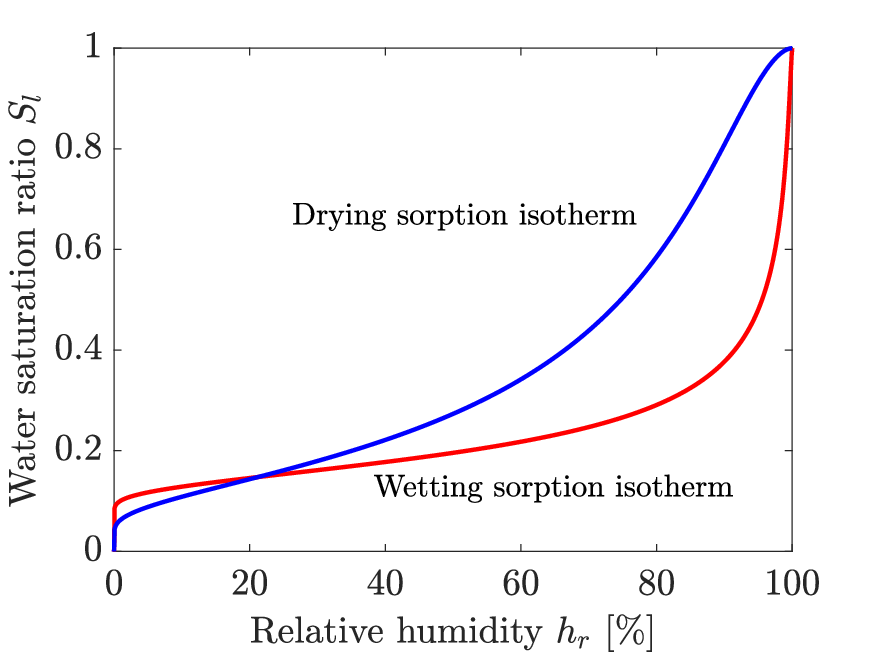}
	\caption{The comparison of wetting and drying sorption isotherm. Curves estimated using Eq. (\ref{Isotherm}) with the experimentally-calibrated values of $\alpha$ and $\beta$ given in Table \ref{tab:tableParam}.} 
	\label{fig:iso_curves}
\end{figure}

For consistency, the same set of concrete material parameters is used across all numerical experiments. The parameters describing water and carbon dioxide transport in concrete are taken from the experimental literature and listed in Table \ref{tab:tableParam}. 

\begin{table}[htb!]
\begin{small}
\begin{longtable}{p{7cm} p{3cm} p{2.5cm} p{2cm}}
\toprule
\textbf{Parameter} & \textbf{Value} & \textbf{Unit} & \textbf{Source} \\
\toprule
\toprule
\multicolumn{4}{ c }{\textbf{$\mathrm{CO}_2$ transport}} \\
\toprule
\toprule
Henry constant $H$ for $\mathrm{CO}_2$ dissolution in water & $ 3.375 \cdot 10^{-4} $ & mol Pa$^{-1}$ m$^{-3}$   & \cite{Papadakis1991} \\
\midrule
Neutralization reaction constant $k_n$ & $ 8.3 $ & m$^3$ mol$^{-1}$ s$^{-1}$ &  \cite{Papadakis1991} \\
\midrule
$\mathrm{OH}^{-}$ equilibrium concentration $c_{\mathrm{OH}^-}^{\mathrm{eq}}$ & $ 43.2 $ & mol m$^{-3}$ & \cite{Papadakis1991} \\
\midrule
Initial $\mathrm{Ca(OH)}_2$ concentration $c_{\mathrm{Ca(OH)}_2}^{0}$ & $1.2 \cdot 10^{-4}$ & mol m$^{-3}$ & \cite{Isgor2004} \\
\midrule
Porosity of fully carbonated concrete $\theta_{c}$  & $ 0.11  $ & - &  \cite{Bretti2022} \\
\bottomrule
\bottomrule
\multicolumn{4}{ c }{\textbf{Water transport}} \\
\toprule
\toprule
Concrete water permeability constant $C$ (wetting and drying) & $ 1.29 \cdot 10^2 $ $\&$ $ 7.4 \cdot 10^6 $ & m$^{4}$ kg$^{-2}$ &  \cite{mainguy2001role,Zhang2022} \\
%Constant $C$ for concrete water permeability (Wetting process) & $ 1.29 \cdot 10^2 $ & m$^{4}$ kg$^{-2}$ &  \cite{Zhang2022} \\
\midrule
Dynamic viscosity of water $\eta$ & $ 10^{-3}  $ & Pa$\cdot$s & \cite{Heider2020} \\
\midrule
Phase-field threshold $\phi_t$ & $ 0.5 $ & - & \cite{Heider2020} \\
\midrule
Density of dried concrete $\rho_{s}$ & $ 2285 $ & kg m$^{-3}$ & \cite{mainguy2001role} \\
\midrule
Sorption isotherm parameter $\alpha$ (wetting and drying) & $ 0.9 \cdot 10^6 $ $\&$ $ 18.62 \cdot 10^6 $ & Pa & \cite{mainguy2001role,Zhang2022} \\
\midrule
Sorption isotherm parameter $\beta$ (wetting and drying) & $ 3.85 $ $\&$ $ 2.27 $ & - & \cite{mainguy2001role,Zhang2022} \\
\toprule
\toprule
\multicolumn{4}{ c }{\textbf{Saturation-dependent corrosion current density model}} \\
\toprule
\toprule
 Maximum effective current density $i_{max}$ & $ 3.7 $ & \unit{\micro\ampere\per\centi\metre^2} & \cite{Stefanoni2019}  \\
\midrule
 Constant $k$ & $ 10^{-3}  $ & - & \cite{Stefanoni2019} \\
\midrule
 Critical porosity $\theta_{crit}$ & $ 0.185  $ & - &  \cite{Stefanoni2019} \\
\toprule
%\multicolumn{4}{ c }{\textbf{{Wetting sorption isotherm}}} \\
%\toprule
%\toprule
%Parameter $\alpha$ & $ 0.9 \cdot 10^6 $ & Pa & \cite{Zhang2022} \\
%\midrule
%Parameter $\beta$ & $ 3.85 $ & - & \cite{Zhang2022} \\
%\end{tabular}
\bottomrule
\caption{Model parameters for describing water and carbon dioxide transport in concrete.}
\label{tab:tableParam}
\end{longtable}
\end{small}
\end{table}

Some parameters deserve detailed consideration. In the proposed model, the water permeability of uncracked concrete $\bm{K}_{m}$ is not constant in time but rather evolves with the changing porosity of concrete caused by carbonation. For this reason, it is important to accurately determine the magnitude of parameter $ C $, which links water permeability and porosity, see Eq. (\ref{isoAbsPerm1}). Here, $ C = 7.4 \cdot 10^6  $ m$^{4}$ kg$^{-2}$ is adopted for the drying process, which leads to permeability values in the order of magnitude of $10^{-21} $ m$^2$, as reported by \citet{mainguy2001role}. The value of $ C = 1.29 \cdot 10^2  $ m$^{4}$ kg$^{-2}$ is adopted for the wetting process, as it leads to permeability values in the order of magnitude of $10^{-16}$ m$^2$, as measured by \citet{Zhang2022}. Thus, water permeability values employed in case studies 1 (Section \ref{Sec:ResValWatTranDry}) and 2 (Section \ref{Sec:ResValWatTranWet}) are consistent with experimentally measured permeabilities reported in their respective studies.
Let us note here that \citet{Zhang2016a} analysed experimentally measured permeability of cementitious materials reported in the literature and concluded that water permeability is on the order of $10^{-21}$ m$^2$ for both pastes or concretes with w/c between 0.4 and 0.5 made from Portland cement. While a permeability reported by \citet{mainguy2001role} (w/c = 0.48, Portland cement) matches this conclusion perfectly, we can see that the permeability recovered by \citet{Zhang2016a} is several orders of magnitude larger. Although the samples of \citet{Zhang2016a} had different composition (w/c = 0.6, blended cement containing burnt oil shale and limestone), \citet{Zhang2016a} concluded that this discrepancy likely results primarily from the carbonation-induced coarsening of pores. Regarding case study 3 (Section \ref{Sec:ResValWatTranCrack}), 
\citet{Michel2018} did not experimentally investigate the permeability of his cracked concrete samples (w/c = 0.5, Portland cement) and thus for this study, it was roughly estimated to be the same as in case study 2 (Section \ref{Sec:ResValWatTranWet}). An excellent match of numerically predicted and experimentally measured extent of water distribution in time suggests that the choice of permeability is appropriate. 

The sorption isotherm constants were experimentally calibrated following the studies by \citet{mainguy2001role} and \citet{Zhang2022} for drying and wetting conditions, respectively. It is worth noting that the authors of these studies recovered $\alpha$ and $\beta$ from experimental data on different concrete samples. This results in both curves crossing at approximately 20$\%$ relative humidity which would theoretically not be expected.  Fig. \ref{fig:iso_curves} shows the differences between the sorption isotherm for wetting and its counterpart for drying. Again, values numerically employed in case studies 1 (Section \ref{Sec:ResValWatTranDry}) and 2 (Section \ref{Sec:ResValWatTranWet}) are consistent with experimentally recovered sorption isotherm constants in these studies. For the sake of the applicability of these values, it is important to mention that \citet{Zhang2022} notes that because the equilibrium water content in a cementitious material is mainly controlled by the amount of dry-hardened cement paste, it can be expected that cementitious materials with the same type of cement and w/c ratio have a very similar sorption isotherm. In the case study 3 (Section \ref{Sec:ResValWatTranCrack}), \citet{Michel2018} did not experimentally measure sorption isotherm. For this reason, the same values of the wetting isotherm constants were used as in case study 2 (Section \ref{Sec:ResValWatTranWet}). While this constitutes an approximation, a parameter sensitivity study reveals only a small influence on the numerical results, with the influence being only noticeable in the water distribution at the very early stages.    

Let us also reiterate that the chosen carbonation model and values of its parameters are aimed at modelling Portland cement-based materials. Although the samples of \citet{Liu2020} (w/c = 0.55) considered in case study 4 (Section \ref{Sec:ResValCarb}) contained fly ash in addition to Portland cement, an excellent fit of predicted and experimentally measured carbonation penetration suggests that the choice of material parameters is appropriate in this case too.    

\FloatBarrier
\subsection{Case study 1: Drying of uncracked concrete}
\label{Sec:ResValWatTranDry}

\begin{figure}[!htb]
    \begin{center}    
    %\begin{adjustbox}{minipage=\linewidth,scale=0.80}
    \begin{subfigure}[!htb]{0.39\textwidth}
    \centering
    \includegraphics[width=\textwidth]{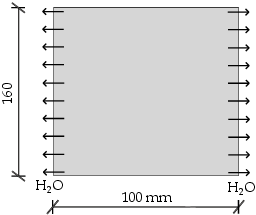}
    \caption{}
    \label{fig:sMG}
    \end{subfigure} 
    \hfill
    \begin{subfigure}[!htb]{0.59\textwidth}
    \centering
    \includegraphics[width=\textwidth]{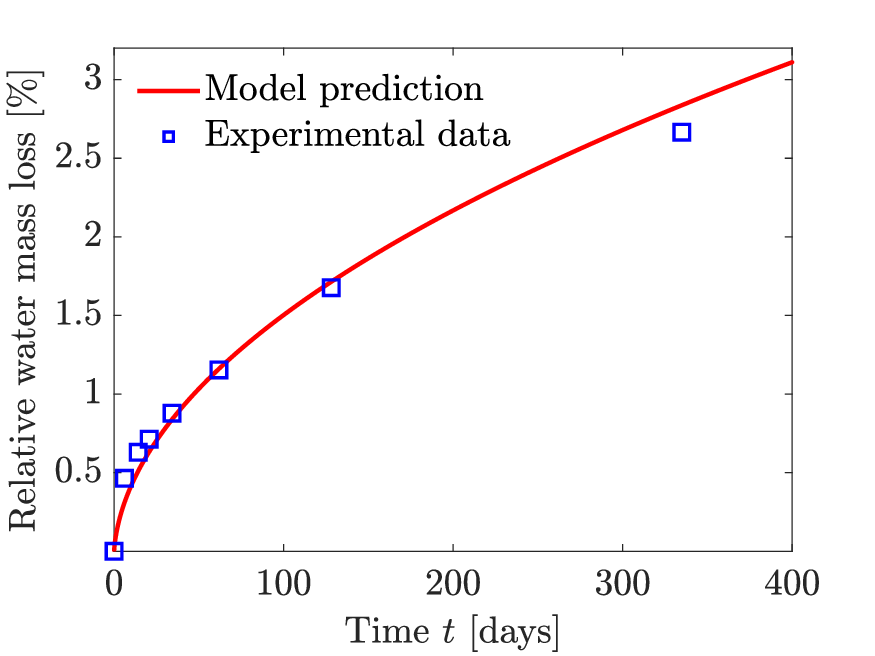}
    \caption{}
    \label{fig:PMD}
    \end{subfigure}    
    \end{center}
    \caption{Simulation of the drying test of \citet{Baroghel-Bouny1999}: (a) geometry of the cross-section of the cylindrical cement paste specimen, and (b) comparison of predicted and experimentally measured water mass loss in time (expressed in percents of the original water mass content).}  
\label{fig:SpResults}
\end{figure}

The ability of the model to predict water transport under drying conditions is evaluated by simulating the experimental tests by \citet{Baroghel-Bouny1999}. Their experiments were based on two-year-old cylindrical cement paste samples with 160 mm diameter and 100 mm height, as shown in the cross-section geometry provided in Fig. \ref{fig:sMG}. Moisture exchange was allowed only on the flat base boundaries while it was prevented on the the curved boundary. During the drying process, the external relative humidity was $50 \%$ and the temperature was $20^{\circ} \mathrm{C}$. Initial relative humidity before drying was measured to be $87 \%$. Because the distribution of initial relative humidity was reported to be uniform, the drying test was simulated as a one-dimensional problem. The porosity of the sample, as measured by \citet{Baroghel-Bouny1999}, was $\theta = 0.12$. The relevant experimental outcome reported by \citet{Baroghel-Bouny1999} is the relative water mass loss, which is numerically estimated as 
\begin{equation}
    \Delta w_r(t) = 100\displaystyle \int_{\Omega_{c}} \frac{S_{l}(0)-S_{l}(t)}{S_{l}(0)} \dd V
\end{equation}
The predicted water mass loss as a function of time is given in Fig. \ref{fig:PMD}, together with the experimental results. A very good agreement between model predictions and experiments is observed.

\FloatBarrier
\subsection{Case study 2: Wetting of uncracked concrete}
\label{Sec:ResValWatTranWet}

\begin{figure}[!htb]
    \begin{center}    
    %\begin{adjustbox}{minipage=\linewidth,scale=0.80}
    \begin{subfigure}[!htb]{0.33\textwidth}
    \centering
    \includegraphics[width=\textwidth]{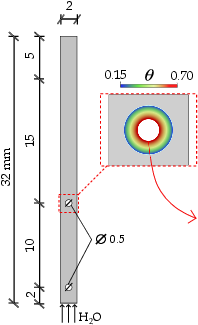}
    \caption{}
    \label{fig:sZ}
    \end{subfigure} 
    \hfill
    \begin{subfigure}[!htb]{0.65\textwidth}
    \centering
    \includegraphics[width=\textwidth]{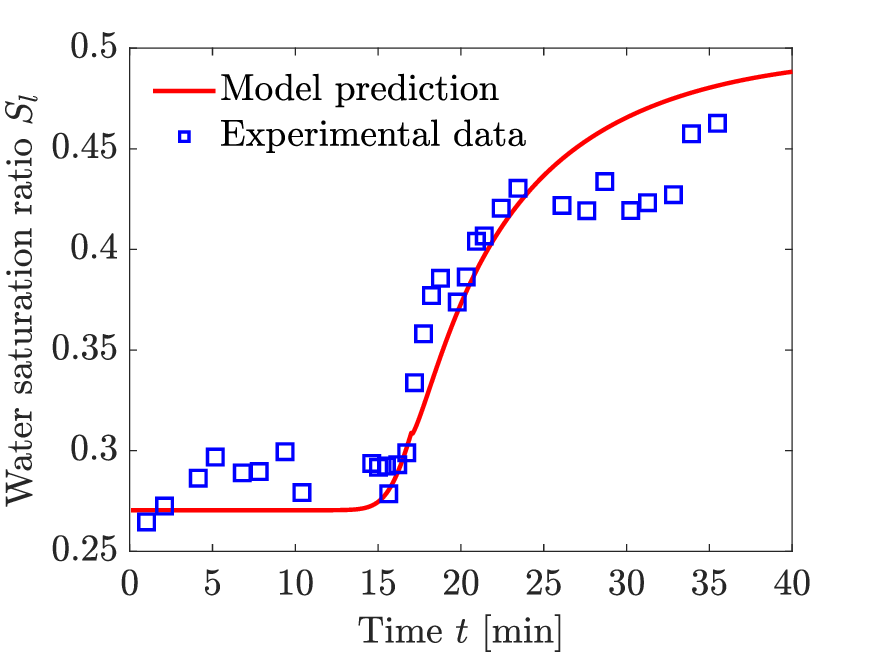}
    \caption{}
    \label{fig:SW}
    \end{subfigure}    
    \end{center}
    \caption{Simulation of the wetting test of \citet{Zhang2022}: (a) Geometry of the cross-section of the mortar specimen with two embedded steel wires, with the inset figure showcasing how the gradual change in porosity reported by \citet{Zhang2022} is accounted for; and (b) comparison of the evolution of the predicted and experimentally measured water saturation ratios in the vicinity of the upper steel wire.}  
\label{fig:SpResults}
\end{figure}

We proceed now to examine the model's capabilities in predicting water transport under wetting conditions. To this end, numerical predictions are compared against the experimental results by \citet{Zhang2022}. \citet{Zhang2022} conducted experiments on a 3 mm thick, 32 mm x 2 mm rectangular mortar sample with two embedded steel wires (see Fig. \ref{fig:sZ}), which was placed on top of a distilled water reservoir such that the lower surface was permanently wetted. Water evaporation from both front and back surfaces was prevented by covering them with aluminium adhesive sheets. Before the wetting test, specimens were carbonated and then stored for seven days in an environment of 53$\%$ relative humidity. Water saturation on the surface of the upper steel wire was monitored during the wetting test. Mortar porosity was measured to be about $\theta = 0.15$ but was reported to increase significantly in the close vicinity of the steel wires. Thus, as suggested by \citet{Zhang2022}, a highly porous layer of 0.3 mm is considered, where porosity changes linearly between 15 and 70$\%$ (see Fig. \ref{fig:sZ}). The simulation and experimental results are given in Fig. \ref{fig:SW}. A very good level of agreement is obtained across the whole time spectrum, with only small differences being observed around the 30-35 min interval, where the scatter in the experimental data appears to be the largest.               

\FloatBarrier
\subsection{Case study 3: Wetting of cracked concrete}
\label{Sec:ResValWatTranCrack}

The presence of cracks significantly affects water transport. To validate the ability of the model to accurately capture the enhancement of water transport through cracks, we choose to model the experiments by \citet{Michel2018}.

\begin{figure}[!htb]
	\centering
	\includegraphics[width=0.5\textwidth]{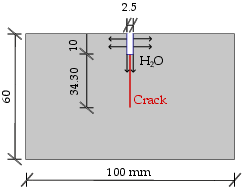}
	\caption{Simulation of the wetting test of \citet{Michel2018} -- cross-section of the concrete samples containing a single crack.} \label{fig:sMP} 
\end{figure}

\begin{figure}[H]
	\centering
	\includegraphics[width=0.8\textwidth]{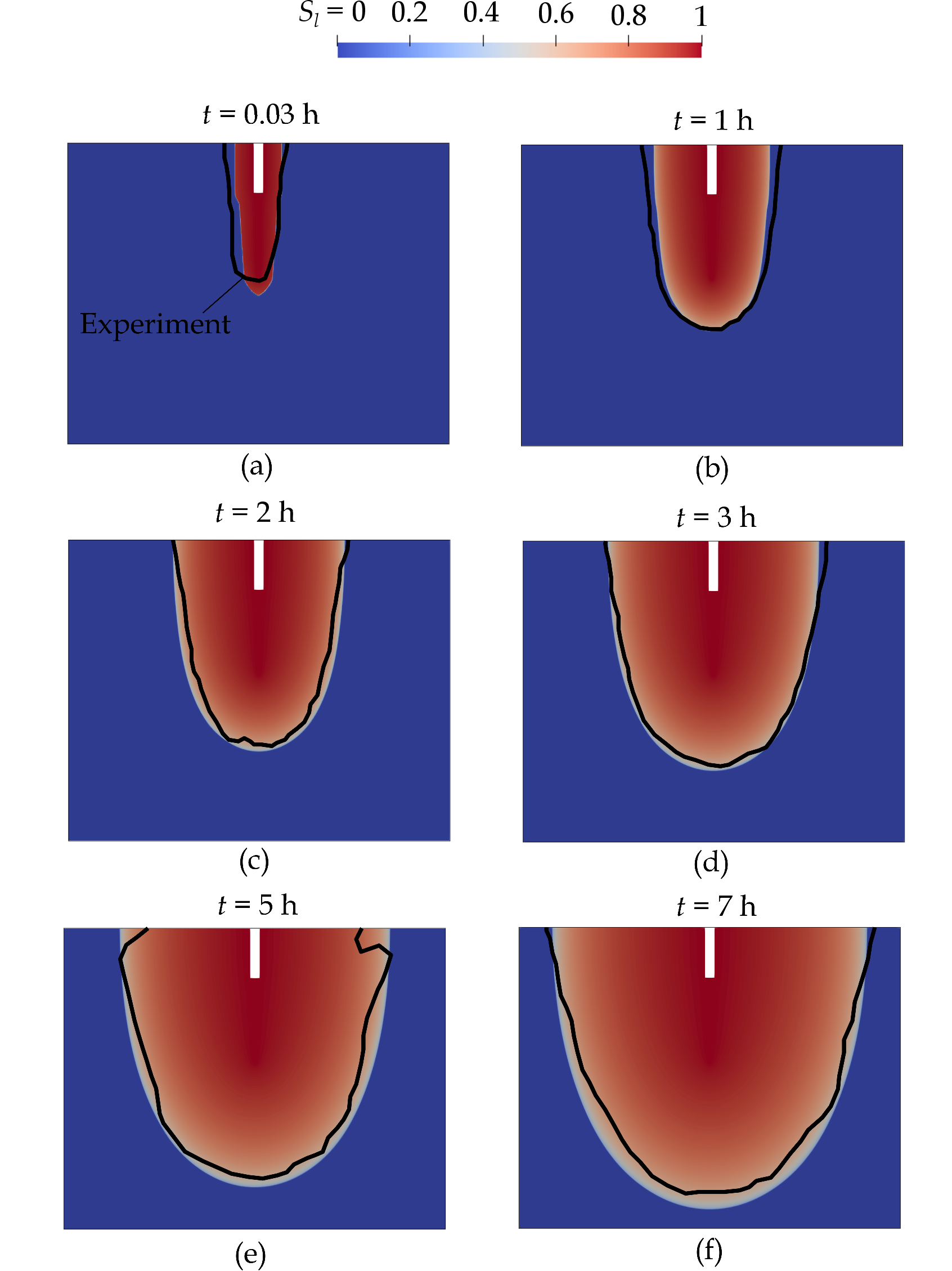}
	\caption{Simulation of the wetting tests of \citet{Michel2018} of concrete samples with a single crack. Contours of simulated saturation ratio and experimentally measured envelope of water distribution (black line) after a time of (a) 0.03 h, (b) 1 h, (c) 2 h, (d) 3 h, (e) 5 h, and (f) 7 h.} 
\label{fig:resMP}
\end{figure}

\citet{Michel2018} used non-destructive X-ray attenuation to monitor the evolution of water saturation in samples containing a single crack. Specifically, as shown in Fig. \ref{fig:sMP}, they employed two-year-old 100x100 mm rectangular samples with a thickness of 50 mm. The sample porosity was estimated from the maximum measured water saturation and reported to be equal to $\theta = 0.12$. \citet{Michel2018} introduced cracks into the notched samples by means of a splitting load, which was applied via a rigid wedge using crack mouth opening displacement (CMOD) control. All specimens were conditioned at $50\%$ relative humidity and a temperature of $20^{\circ} \mathrm{C}$ for at least 1 year before wetting. The simulated part of the domain with the crack is depicted in Fig. \ref{fig:sMP}. The crack was measured to be 34.3 mm long and 0.043 mm wide, with the crack opening decreasing with depth, as reported by \citet{Michel2018}. Accordingly, we introduce the crack into the model by assigning $\phi=1$ in the centre of the crack and smearing the cracked domain using Eq. (\ref{pfEq}). %A linear variation of the phase field length scale is adopted, going from $\ell = 0.043$ mm on the concrete surface to $ \ell = 0$ at the crack tip. 
The choice of $ \ell = 0.043$ mm leads to the experimentally reported opening of the crack on the surface such that $\phi \geq \phi_t $ on the region of the same width. During the wetting test, a cast-in recess and a cut notch above the crack were used as a reservoir of liquid water. For this reason, $ S_l = 1 $ was considered on the boundaries of the notch. On the top surface of the specimen, water saturation equivalent to $65\%$ relative humidity is considered, as these are the conditions relevant to the X-ray chamber. Mimicking the experimental setup, we consider zero flux on the remaining concrete surfaces. In Fig. \ref{fig:resMP}, the experimentally measured envelope of a partially water-saturated region is compared with the predicted distribution of water saturation at different times, up to seven hours from the beginning of the wetting process.

The notable impact of cracks in water transport is evident from the very early stages of the test, see Fig. \ref{fig:resMP}a ($t=0.03$ h). Water transport is significantly enhanced in the vertical direction, particularly at the beginning of the experiment. With time, the water distribution is more uniform and the contour shows a semi-ellipsoidal shape, with both axes of the ellipse expanding in time. For all the evaluated times, the extent of water distribution agrees with their experimental counterparts very well, indicating that the model is able to accurately predict water transport in cracked concrete samples undergoing wetting.  

\FloatBarrier
\subsection{Case study 4: Carbonation of variably water-saturated concrete}
\label{Sec:ResValCarb}

\begin{figure}[!htb]
    \begin{center}    
    %\begin{adjustbox}{minipage=\linewidth,scale=0.80}
    \begin{subfigure}[!htb]{0.35\textwidth}
    \centering
    \includegraphics[width=\textwidth]{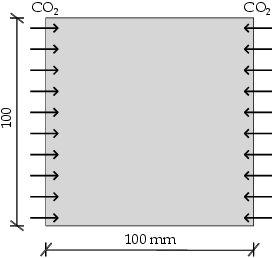}
    \caption{}
    \label{fig:sL}    
    \end{subfigure} 
    \hfill
    \begin{subfigure}[!htb]{0.63\textwidth}
    \centering
    \includegraphics[width=\textwidth]{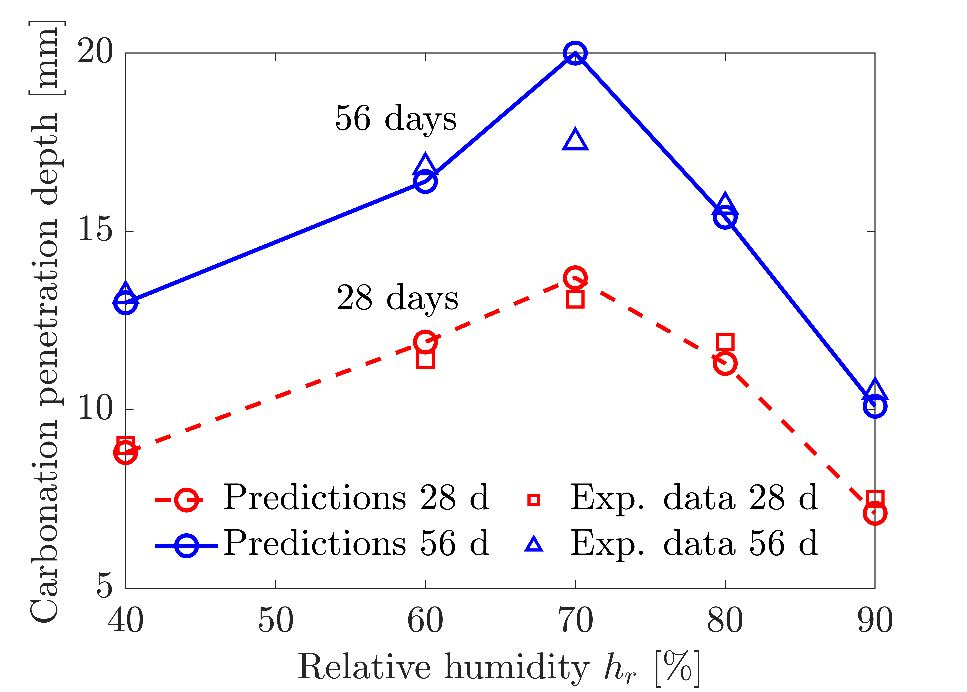}
    \caption{}
    \label{fig:cB}    
    \end{subfigure}    
    \end{center}
    \caption{Simulation of the carbonation tests of \citet{Liu2020}: (a) Geometry of the cross-section of the concrete specimen, and (b) comparison of predicted and experimentally measured carbonation depth for varying relative humidity. The proposed model accurately predicts the concave dependency of the carbonation depth on the relative humidity. This results from the competition between CO$_2$ diffusivity decreasing with increasing water saturation and the reaction rate of neutralization reaction increasing with increasing water saturation.}  
\label{fig:carbVal}
\end{figure}

We shall now use the carbonation depth measurements by \citet{Liu2020} to investigate the ability of the model to capture the interplay between carbonation and water saturation. \citet{Liu2020} conducted experiments on concrete cubes of characteristic length 100 mm that had been cured for 28 days and subsequently dried for 2 days at $60^{\circ} \mathrm{C}$, so as to minimise the presence of water. The initial porosity is deemed to be $\theta_{0} = 0.26$, based on the model of \citet{Powers1946} for a reported water-cement ratio of 0.55 and assuming a degree of hydration of 0.9. As sketched in Fig. \ref{fig:sL}, the experiments involved using an environmental chamber to expose two opposite sides of the sample to a constant temperature of $19.85^{\circ} \mathrm{C}$ and a fixed carbon dioxide concentration of 20$\%$ $\mathrm{CO}_2$. In contrast, the relative humidity was varied from 40 to 90\%, so as to estimate carbon penetration depths as a function of time and relative humidity $h_r$. As shown in Fig. \ref{fig:cB}, measurements were taken after 28 and 56 days of exposure.  

Model predictions are also shown in Fig. \ref{fig:cB}, using circle symbols and lines. 
Numerical predictions for both 28 and 56 days agree with experimental measurements very well. Only for the case of 70$\%$ relative humidity and 56 days, the model overestimates the carbonation depth, by approximately 2 mm. In Fig. \ref{fig:cB} we can see that the curve depicting the dependence of carbonation depth on relative humidity is concave. This is caused by two competing effects arising with increasing water saturation -- an increase of the reaction rate of the neutralization reaction, see Eq. (\ref{eq:CO2_diffusion}), and a decrease of carbon dioxide diffusivity, see Eq. (\ref{eq:Diff_coeff_CO2}). If water saturation is low, gaseous carbon dioxide penetrates the concrete pore space easily, but the carbonation reaction takes place in pore solution and is thus hindered by low water content. On the other hand, if water saturation of pore space is high, carbonation proceeds quickly but the transport of carbon dioxide is significantly hindered. As shown in Fig. \ref{fig:cB}, this leads to the existence of an optimal water saturation point under which the quickest carbonation rate is observed. For the described tests of \citet{Liu2020}, the maximum carbonation depth was observed for a 70$\%$ relative humidity, and this is also the case in our simulations.

\FloatBarrier
\subsection{Case study 5: Insights into the interplay between carbonation, cyclic wetting/drying and corrosion}
\label{Sec:ResCarbCorrDenWetDry}

\begin{figure}[!htb]
	\centering
	\includegraphics[width=0.9\textwidth]{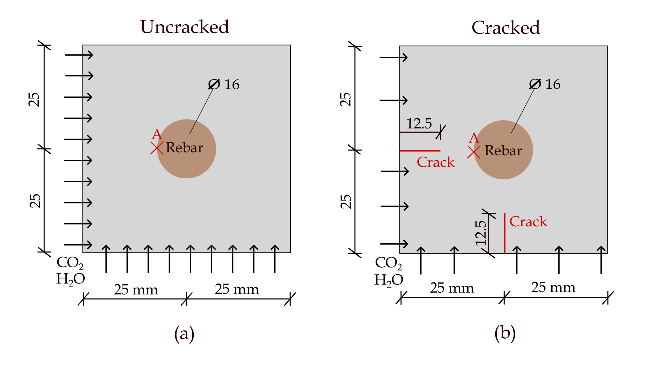}
	\caption{Geometry of the cross-section of reinforced concrete samples subjected to carbon dioxide penetration and cyclic wetting and drying: (a) uncracked sample and (b) cracked sample. Point A marks the location where the corrosion current density is evaluated.} 
\label{fig:schemeWetDry} 
\end{figure}

\begin{figure}[H]
	\centering
	\includegraphics[width=0.8\textwidth]{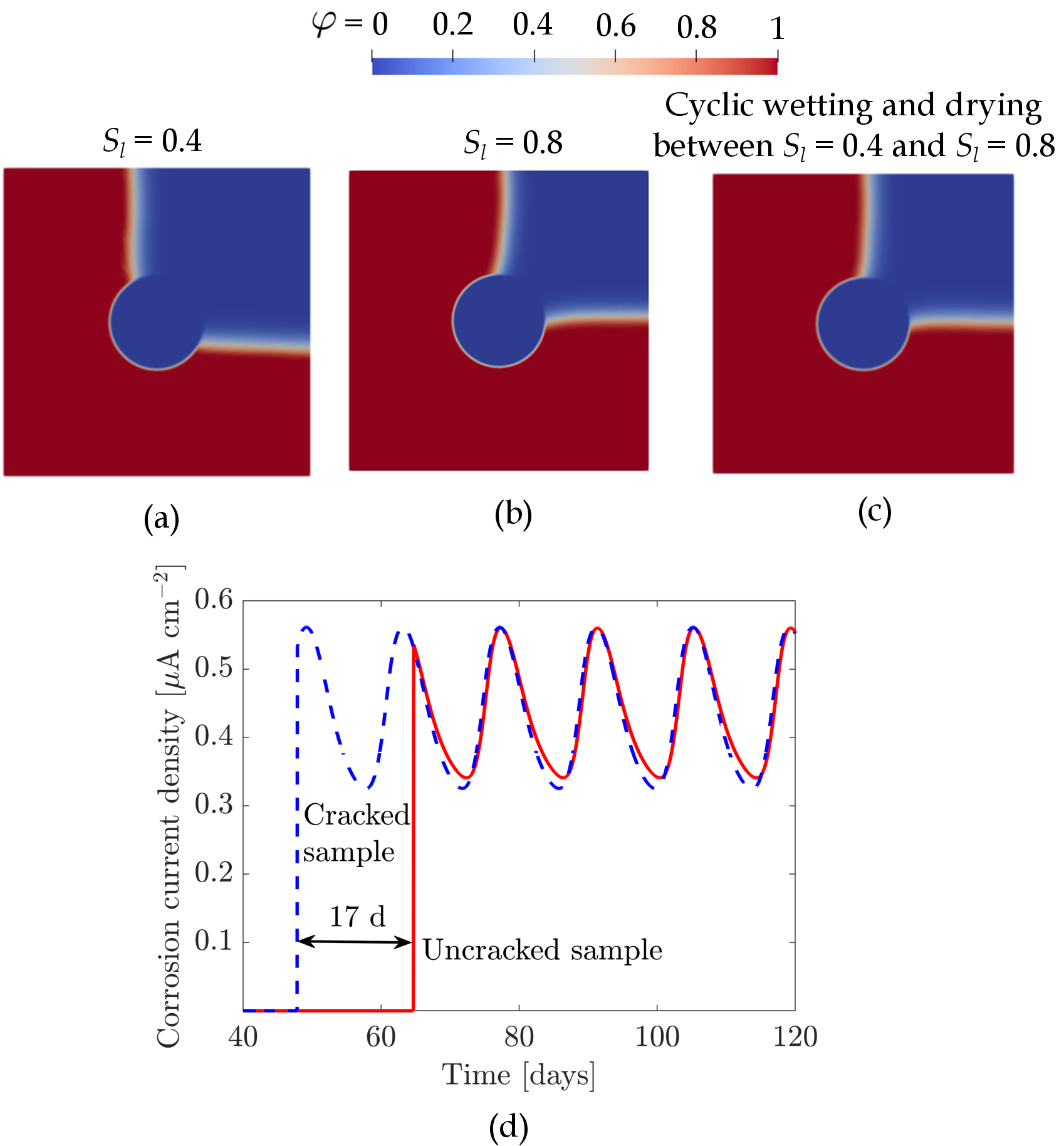}
	\caption{Carbonation front after 120 days for an uncracked concrete sample subjected to (a) constant 40$\%$ water saturation of concrete pore space, (b) constant 80$\%$ water saturation, and (c) initial 40$\%$ water saturation followed by cyclic 40 - 80$\%$ water saturation of the exposed boundaries. In (d), the evolution of corrosion current density under cyclic water saturation at point A (see Fig. \ref{fig:schemeWetDry}) is compared for cracked and uncracked samples.} 
\label{fig:varSpCarb} 
\end{figure}

During the carbonation process, exposed reinforced concrete structures are subjected to cyclic wetting and drying. Because the carbonation rate is highly dependent on the water saturation of concrete pores, as showcased in the previous section, the time to steel depassivation is also sensitive to these humidity changes. Also, changes in water saturation lead to changes in corrosion current density (see Section \ref{subSec:CorrCurrMod}). Cracks resulting from loading, temperature gradients, shrinkage and other effects are commonly present in reinforced concrete structures and significantly enhance the transport of water and carbon dioxide, accelerating carbonation. To demonstrate how the proposed model can be used to investigate the interaction of these processes during alternating wetting and drying, the behaviour of a 50 mm x 50 mm concrete sample reinforced with a single rebar of 16 mm diameter is investigated (see Figs. \ref{fig:varSpCarb}a and \ref{fig:varSpCarb}b). A 16$\%$ porosity was considered. Penetration of water and carbon dioxide was allowed on two perpendicular surfaces while zero flux for both species was considered on the two remaining surfaces. The specimen had an initial pore water saturation of 40$\%$  and the saturation of exposed boundaries changed periodically from 40$\%$ to 80$\%$. The exposed boundary saturation varied in time as $ S_l = 0.4\left(1 + 0.5(\text{sin}(\pi t/7 + 1.5 \pi) + 1)\right)$ and the cycle time was thus 14 days. For the sake of simplicity, only the wetting isotherm is employed. To investigate the effect of cracks, both uncracked (Fig. \ref{fig:varSpCarb}a) and cracked (Fig. \ref{fig:varSpCarb}b) samples are considered. The cracked sample contains two 15 mm long cracks, which start from each of the exposed surfaces. These cracks were numerically introduced by prescribing $ \phi = 1 $ in the centre of the crack and then regularised using Eq. (\ref{pfEq}) and adopting a phase field length scale of $ \ell = 0.5$ mm.

Contours of the carbonation front variable are shown in Figs. \ref{fig:varSpCarb}a-c for selected concrete water saturation levels after 120 days of exposure. The results obtained for constant water saturation values of 40$\%$ and 80$\%$ are given in Figs. \ref{fig:varSpCarb}a and \ref{fig:varSpCarb}b, respectively. It is useful to compare them with the results obtained for variable exposed boundary water saturation, Fig. \ref{fig:varSpCarb}c. While constant 40$\%$ water saturation leads to the slowest advance of carbonation front, in agreement with expectations, periodic wetting and drying between 40$\%$ and 80$\%$ leads to nearly the same result as a constant 80$\%$ saturation. This is because periodic wetting events are able to sustain high water saturation which allows for the quicker advance of the carbonation front. Thus, our results highlight how short wetting events can have a profound impact on the long-term water saturation of concrete. This was documented for example by \citet{Andrade1999}, who found out that rain periods are the main factor influencing the internal relative humidity of unsheltered concrete samples. For this reason, even short cyclic periods of high external humidity seem to potentially lead to nearly the same rebar depassivation as if the maximum of periodic boundary saturation remained on the whole domain during the entire time period. Also, it has been observed that the average corrosion current density in samples subjected to cyclic wetting and drying could be comparable to permanently wet samples \cite{AnguloRamirez2023}.\\ 

The combined effect of variable moisture saturation and cracks on the variation of the corrosion current density is shown in Fig. \ref{fig:varSpCarb}d. As it can be observed, the corrosion current density (evaluated at point A, see Fig. \ref{fig:schemeWetDry}), exhibits significant sensitivity to water concentration, varying between 0.28 and 0.56 \unit{\micro\ampere\per\centi\metre^2}. This finding emphasises the need to account for the role of water saturation on carbonation-induced corrosion current density. Also, it can be observed that the presence of cracks significantly shortens the time to corrosion initiation, going from approximately 65 to 48 days. Thus, durability models neglecting the role of cracks in enhancing transport and corrosion should be revisited.  

\FloatBarrier
\section{Conclusions}
\label{Sec:Conclusions}

In this study, we have presented a novel coupled model for water transport, carbonation and corrosion in concrete. Additionally, the new formulation presented can capture the role of cracks in enhancing water and CO$_2$ transport. The proposed model was extensively validated against experimental data for wetting \cite{Zhang2022} and drying \cite{Baroghel-Bouny1999} of uncracked concrete, wetting of cracked concrete \cite{Michel2018}, and carbonation under varying water saturation \cite{Liu2020}. Although all these studies considered different concrete samples, their main difference from the water transport and carbonation perspective lies in the porosity of cement paste and the structure of pore space. Porosity is the principal variable of the model and when set according to measured data or estimated when no measurements were available, the model proved to be able to replicate experimentally measured results from \cite{Zhang2022,Baroghel-Bouny1999,Michel2018,Liu2020} very well. However, let us note here that carbonation was found to affect the structure of pore space \cite{Zhang2022} and although data from this study allowed for the validation of the wetting of uncracked carbonated concrete, other water transport regimes such as drying or wetting of cracked concrete were possible to validate only with uncarbonated samples because of the lack of available experimental data. If such data will become available in the future, we recommend repeated model validation.
Furthermore, we build upon the abilities of the model to gain insight into the interplay between corrosion, cracks, and water and CO$_2$ transport under cyclic wetting and drying conditions. Key findings include:
\begin{itemize}
\item The model can accurately simulate the wetting and drying of concrete under isothermal conditions, including the enhancement of water transport through cracks. 
\item The model is able to accurately capture the impact of water saturation on carbonation, including the interplay between the opposite trends of saturation-dependent neutralization reaction rate and saturation-dependent carbon dioxide diffusivity. An optimal water saturation point is identified for an intermediate value of humidity. 
\item Cyclic wetting and drying leads to significant acceleration in the evolution of the carbonation front. Because the drying process tends to be much slower than wetting, even a short intense wetting period can significantly accelerate the carbonation process. 
\item The corrosion current density changes significantly with varying concrete saturation such that for a concrete specimen with boundary water saturation periodically varying between 40 and 80$\%$, the corrosion current density periodically drops to 56$\%$ of its maximum value.   
\item The time to corrosion initiation is significantly shortened if surface cracks are present, a 30$\%$ reduction is observed in the investigated case study.  
\end{itemize}

The model can readily be extended to account for the role of: (i) growing cracks, e.g. by incorporating a phase field evolution law \cite{Freddi2022,Korec2023}, (ii) sorption hysteresis effects, e.g., as in \citet{Zhang2015a}, (iii) chloride-induced corrosion, to capture the interplay between carbonation- and chloride-driven corrosion processes \cite{Zhu2016a}, (iv) types of wetting cycles, and (v) the role of oxygen saturation, of potentially high importance yet still not completely understood  \cite{Andrade2023a,Stefanoni2019,hageman2023corrosion}. 

\section{Acknowledgements}
\label{Acknowledge of funding}

The authors would like to express their gratitude to Prof Milan Jirásek (Czech Technical University in Prague) for his invaluable advice on the formulation of the water transport model and to Prof Hong S. Wong (Imperial College London) for stimulating discussions. E. Korec acknowledges financial support from the Imperial College President’s PhD Scholarships. E. Mart\'{\i}nez-Pa\~neda was supported by a UKRI Future Leaders Fellowship [grant MR/V024124/1]. F. Freddi and L. Mingazzi were supported by the Project ECOSISTER SPOKE 4 funded under the National Recovery and Resilience Plan (NRRP), Mission 04 Component 2 Investment 1.5 – NextGenerationEU, Call for tender n. 3277 dated 30/12/2021.


\bibliographystyle{elsarticle-num-names}
%\bibliography{../../mend/moistureTranport.bib}

\begin{thebibliography}{71}
\expandafter\ifx\csname natexlab\endcsname\relax\def\natexlab#1{#1}\fi
\providecommand{\url}[1]{\texttt{#1}}
\providecommand{\href}[2]{#2}
\providecommand{\path}[1]{#1}
\providecommand{\DOIprefix}{doi:}
\providecommand{\ArXivprefix}{arXiv:}
\providecommand{\URLprefix}{URL: }
\providecommand{\Pubmedprefix}{pmid:}
\providecommand{\doi}[1]{\href{http://dx.doi.org/#1}{\path{#1}}}
\providecommand{\Pubmed}[1]{\href{pmid:#1}{\path{#1}}}
\providecommand{\bibinfo}[2]{#2}
\ifx\xfnm\relax \def\xfnm[#1]{\unskip,\space#1}\fi
%Type = Incollection
\bibitem[{Poursaee et~al.(2016)Poursaee, K., Lasa, Hansson, McDonald, Yeomans,
  Holland, Kurtis, Kahn, Moriconi, Ziehl, ElBatanouny, {Burkan Isgor}, Andrade,
  Spragg, Qiao, and Weiss}]{Poursaee2016a}
\bibinfo{author}{A.~Poursaee}, \bibinfo{author}{L.~K.},
  \bibinfo{author}{I.~Lasa}, \bibinfo{author}{C.~Hansson},
  \bibinfo{author}{D.~McDonald}, \bibinfo{author}{S.~Yeomans},
  \bibinfo{author}{R.~Holland}, \bibinfo{author}{K.~Kurtis},
  \bibinfo{author}{L.~Kahn}, \bibinfo{author}{G.~Moriconi},
  \bibinfo{author}{P.~Ziehl}, \bibinfo{author}{M.~ElBatanouny},
  \bibinfo{author}{O.~{Burkan Isgor}}, \bibinfo{author}{C.~Andrade},
  \bibinfo{author}{R.~Spragg}, \bibinfo{author}{C.~Qiao},
  \bibinfo{author}{J.~Weiss},
\newblock \bibinfo{title}{{Corrosion of Steel in Concrete Structures}},
\newblock in: \bibinfo{editor}{A.~Poursaee} (Ed.),
  \bibinfo{booktitle}{Corrosion of Steel in Concrete Structures},
  \bibinfo{publisher}{Woodhead Publishing}, \bibinfo{address}{Oxford},
  \bibinfo{year}{2016}, pp. \bibinfo{pages}{19--33}.
%Type = Article
\bibitem[{Angst(2018)}]{Angst2018a}
\bibinfo{author}{U.~M. Angst},
\newblock \bibinfo{title}{{Challenges and opportunities in corrosion of steel
  in concrete}},
\newblock \bibinfo{journal}{Materials and Structures} \bibinfo{volume}{51}
  (\bibinfo{year}{2018}) \bibinfo{pages}{1--20}.
%Type = Article
\bibitem[{Andrade(2019)}]{Andrade2019}
\bibinfo{author}{C.~Andrade},
\newblock \bibinfo{title}{{Propagation of reinforcement corrosion: principles,
  testing and modelling}},
\newblock \bibinfo{journal}{Materials and Structures} \bibinfo{volume}{52}
  (\bibinfo{year}{2019}) \bibinfo{pages}{1--26}.
%Type = Book
\bibitem[{Gehlen et~al.(2011)Gehlen, Andrade, Bartholomew, Cairns, Gulikers,
  {Javier Leon}, Matthews, McKenna, Osterminski, Paeglitis, and
  Straub}]{Gehlen2011-za}
\bibinfo{author}{C.~Gehlen}, \bibinfo{author}{C.~Andrade},
  \bibinfo{author}{M.~Bartholomew}, \bibinfo{author}{J.~Cairns},
  \bibinfo{author}{J.~Gulikers}, \bibinfo{author}{F.~{Javier Leon}},
  \bibinfo{author}{S.~Matthews}, \bibinfo{author}{P.~McKenna},
  \bibinfo{author}{K.~Osterminski}, \bibinfo{author}{A.~Paeglitis},
  \bibinfo{author}{D.~Straub}, \bibinfo{title}{{Fib Bulletin 59 - Condition
  control and assessment of reinforced concrete structures exposed to corrosive
  environments (carbonation/chlorides)}}, \bibinfo{publisher}{fib - The
  International Federation for Structural Concrete}, \bibinfo{year}{2011}.
%Type = Book
\bibitem[{Jones and Marsh(1997)}]{British_Cement_Association_BCA1997-jj}
\bibinfo{author}{A.~E.~K. Jones}, \bibinfo{author}{B.~K. Marsh},
  \bibinfo{title}{{Development of an holistic approach to ensure the durability
  of new concrete construction}}, \bibinfo{publisher}{British Cement
  Association (BCA)}, \bibinfo{year}{1997}.
%Type = Article
\bibitem[{Angst et~al.(2020)Angst, Moro, Geiker, Kessler, Beushausen, Andrade,
  Lahdensivu, K{\"{o}}li{\"{o}}, Imamoto, von Greve-Dierfeld, and
  Serdar}]{Angst2020a}
\bibinfo{author}{U.~Angst}, \bibinfo{author}{F.~Moro},
  \bibinfo{author}{M.~Geiker}, \bibinfo{author}{S.~Kessler},
  \bibinfo{author}{H.~Beushausen}, \bibinfo{author}{C.~Andrade},
  \bibinfo{author}{J.~Lahdensivu}, \bibinfo{author}{A.~K{\"{o}}li{\"{o}}},
  \bibinfo{author}{K.~I. Imamoto}, \bibinfo{author}{S.~von Greve-Dierfeld},
  \bibinfo{author}{M.~Serdar},
\newblock \bibinfo{title}{{Corrosion of steel in carbonated concrete:
  Mechanisms, practical experience, and research priorities – A critical
  review by RILEM TC 281-CCC}},
\newblock \bibinfo{journal}{RILEM Technical Letters} \bibinfo{volume}{5}
  (\bibinfo{year}{2020}) \bibinfo{pages}{85--100}.
%Type = Article
\bibitem[{Yang et~al.(2022)Yang, Yang, Caggiano, Ukrainczyk, and
  Koenders}]{Yang2022}
\bibinfo{author}{S.~Yang}, \bibinfo{author}{Y.~Yang},
  \bibinfo{author}{A.~Caggiano}, \bibinfo{author}{N.~Ukrainczyk},
  \bibinfo{author}{E.~Koenders},
\newblock \bibinfo{title}{{A phase-field approach for portlandite carbonation
  and application to self-healing cementitious materials}},
\newblock \bibinfo{journal}{Materials and Structures} \bibinfo{volume}{55}
  (\bibinfo{year}{2022}) \bibinfo{pages}{1--19}.
%Type = Article
\bibitem[{Zhang et~al.(2021)Zhang, Ghouleh, Azar, and Shao}]{Zhang2021a}
\bibinfo{author}{S.~Zhang}, \bibinfo{author}{Z.~Ghouleh},
  \bibinfo{author}{A.~Azar}, \bibinfo{author}{Y.~Shao},
\newblock \bibinfo{title}{{Improving concrete resistance to low temperature
  sulfate attack through carbonation curing}},
\newblock \bibinfo{journal}{Materials and Structures} \bibinfo{volume}{54}
  (\bibinfo{year}{2021}) \bibinfo{pages}{1--18}.
%Type = Article
\bibitem[{Dhandapani et~al.(2022)Dhandapani, Joseph, Bishnoi, Kunther,
  Kanavaris, Kim, Irassar, Castel, Zunino, Machner, Talakokula, Thienel,
  Wilson, Elsen, Martirena, and Santhanam}]{Dhandapani2022}
\bibinfo{author}{Y.~Dhandapani}, \bibinfo{author}{S.~Joseph},
  \bibinfo{author}{S.~Bishnoi}, \bibinfo{author}{W.~Kunther},
  \bibinfo{author}{F.~Kanavaris}, \bibinfo{author}{T.~Kim},
  \bibinfo{author}{E.~Irassar}, \bibinfo{author}{A.~Castel},
  \bibinfo{author}{F.~Zunino}, \bibinfo{author}{A.~Machner},
  \bibinfo{author}{V.~Talakokula}, \bibinfo{author}{K.~C. Thienel},
  \bibinfo{author}{W.~Wilson}, \bibinfo{author}{J.~Elsen},
  \bibinfo{author}{F.~Martirena}, \bibinfo{author}{M.~Santhanam},
\newblock \bibinfo{title}{{Durability performance of binary and ternary blended
  cementitious systems with calcined clay: a RILEM TC 282 CCL review}},
\newblock \bibinfo{journal}{Materials and Structures} \bibinfo{volume}{55}
  (\bibinfo{year}{2022}).
%Type = Article
\bibitem[{von Greve-Dierfeld et~al.(2020)von Greve-Dierfeld, Lothenbach,
  Vollpracht, Wu, Huet, Andrade, Medina, Thiel, Gruyaert, Vanoutrive,
  {Sa{\'{e}}z del Bosque}, Ignjatovic, Elsen, Provis, Scrivener, Thienel,
  Sideris, Zajac, Alderete, Cizer, {Van den Heede}, Hooton, Kamali-Bernard,
  Bernal, Zhao, Shi, and {De Belie}}]{VonGreve-Dierfeld2020}
\bibinfo{author}{S.~von Greve-Dierfeld}, \bibinfo{author}{B.~Lothenbach},
  \bibinfo{author}{A.~Vollpracht}, \bibinfo{author}{B.~Wu},
  \bibinfo{author}{B.~Huet}, \bibinfo{author}{C.~Andrade},
  \bibinfo{author}{C.~Medina}, \bibinfo{author}{C.~Thiel},
  \bibinfo{author}{E.~Gruyaert}, \bibinfo{author}{H.~Vanoutrive},
  \bibinfo{author}{I.~F. {Sa{\'{e}}z del Bosque}},
  \bibinfo{author}{I.~Ignjatovic}, \bibinfo{author}{J.~Elsen},
  \bibinfo{author}{J.~L. Provis}, \bibinfo{author}{K.~Scrivener},
  \bibinfo{author}{K.~C. Thienel}, \bibinfo{author}{K.~Sideris},
  \bibinfo{author}{M.~Zajac}, \bibinfo{author}{N.~Alderete},
  \bibinfo{author}{{\"{O}}.~Cizer}, \bibinfo{author}{P.~{Van den Heede}},
  \bibinfo{author}{R.~D. Hooton}, \bibinfo{author}{S.~Kamali-Bernard},
  \bibinfo{author}{S.~A. Bernal}, \bibinfo{author}{Z.~Zhao},
  \bibinfo{author}{Z.~Shi}, \bibinfo{author}{N.~{De Belie}},
\newblock \bibinfo{title}{{Understanding the carbonation of concrete with
  supplementary cementitious materials: a critical review by RILEM TC
  281-CCC}},
\newblock \bibinfo{journal}{Materials and Structures} \bibinfo{volume}{53}
  (\bibinfo{year}{2020}).
%Type = Article
\bibitem[{Leemann et~al.(2018)Leemann, Pahlke, Loser, and
  Winnefeld}]{Leemann2018}
\bibinfo{author}{A.~Leemann}, \bibinfo{author}{H.~Pahlke},
  \bibinfo{author}{R.~Loser}, \bibinfo{author}{F.~Winnefeld},
\newblock \bibinfo{title}{{Carbonation resistance of mortar produced with
  alternative cements}},
\newblock \bibinfo{journal}{Materials and Structures} \bibinfo{volume}{51}
  (\bibinfo{year}{2018}) \bibinfo{pages}{1--12}.
%Type = Article
\bibitem[{Leemann et~al.(2015)Leemann, Nygaard, Kaufmann, and
  Loser}]{Leemann2015a}
\bibinfo{author}{A.~Leemann}, \bibinfo{author}{P.~Nygaard},
  \bibinfo{author}{J.~Kaufmann}, \bibinfo{author}{R.~Loser},
\newblock \bibinfo{title}{{Relation between carbonation resistance, mix design
  and exposure of mortar and concrete}},
\newblock \bibinfo{journal}{Cement and Concrete Composites}
  \bibinfo{volume}{62} (\bibinfo{year}{2015}) \bibinfo{pages}{33--43}.
%Type = Article
\bibitem[{Auroy et~al.(2015)Auroy, Poyet, {Le Bescop}, Torrenti, Charpentier,
  Moskura, and Bourbon}]{Auroy2015}
\bibinfo{author}{M.~Auroy}, \bibinfo{author}{S.~Poyet}, \bibinfo{author}{P.~{Le
  Bescop}}, \bibinfo{author}{J.~M. Torrenti}, \bibinfo{author}{T.~Charpentier},
  \bibinfo{author}{M.~Moskura}, \bibinfo{author}{X.~Bourbon},
\newblock \bibinfo{title}{{Impact of carbonation on unsaturated water transport
  properties of cement-based materials}},
\newblock \bibinfo{journal}{Cement and Concrete Research} \bibinfo{volume}{74}
  (\bibinfo{year}{2015}) \bibinfo{pages}{44--58}.
%Type = Article
\bibitem[{Papadakis et~al.(1991)Papadakis, Vayenas, and Fardis}]{Papadakis1991}
\bibinfo{author}{V.~G. Papadakis}, \bibinfo{author}{C.~G. Vayenas},
  \bibinfo{author}{M.~N. Fardis},
\newblock \bibinfo{title}{{Fundamental modeling and experimental investigation
  of concrete carbonation}},
\newblock \bibinfo{journal}{ACI Materials Journal} \bibinfo{volume}{88}
  (\bibinfo{year}{1991}) \bibinfo{pages}{363--373}.
%Type = Article
\bibitem[{Saetta et~al.(1993)Saetta, Schrefler, and Vitaliani}]{Saetta1993a}
\bibinfo{author}{A.~V. Saetta}, \bibinfo{author}{B.~A. Schrefler},
  \bibinfo{author}{R.~V. Vitaliani},
\newblock \bibinfo{title}{{The carbonation of concrete and the mechanism of
  moisture, heat and carbon dioxide flow through porous materials}},
\newblock \bibinfo{journal}{Cement and Concrete Research} \bibinfo{volume}{23}
  (\bibinfo{year}{1993}) \bibinfo{pages}{761--772}.
%Type = Article
\bibitem[{Steffens et~al.(2002)Steffens, Dinkler, and Ahrens}]{Steffens2002}
\bibinfo{author}{A.~Steffens}, \bibinfo{author}{D.~Dinkler},
  \bibinfo{author}{H.~Ahrens},
\newblock \bibinfo{title}{{Modeling carbonation for corrosion risk prediction
  of concrete structures}},
\newblock \bibinfo{journal}{Cement and Concrete Research} \bibinfo{volume}{32}
  (\bibinfo{year}{2002}) \bibinfo{pages}{935--941}.
%Type = Article
\bibitem[{Isgor and Razaqpur(2004)}]{Isgor2004}
\bibinfo{author}{O.~B. Isgor}, \bibinfo{author}{A.~G. Razaqpur},
\newblock \bibinfo{title}{{Finite element modeling of coupled heat transfer,
  moisture transport and carbonation processes in concrete structures}},
\newblock \bibinfo{journal}{Cement and Concrete Composites}
  \bibinfo{volume}{26} (\bibinfo{year}{2004}) \bibinfo{pages}{57--73}.
%Type = Article
\bibitem[{Song et~al.(2006)Song, Kwon, Byun, and Park}]{Song2006}
\bibinfo{author}{H.~W. Song}, \bibinfo{author}{S.~J. Kwon},
  \bibinfo{author}{K.~J. Byun}, \bibinfo{author}{C.~K. Park},
\newblock \bibinfo{title}{{Predicting carbonation in early-aged cracked
  concrete}},
\newblock \bibinfo{journal}{Cement and Concrete Research} \bibinfo{volume}{36}
  (\bibinfo{year}{2006}) \bibinfo{pages}{979--989}.
%Type = Article
\bibitem[{Bary and Sellier(2004)}]{Bary2004}
\bibinfo{author}{B.~Bary}, \bibinfo{author}{A.~Sellier},
\newblock \bibinfo{title}{{Coupled moisture - Carbon dioxide-calcium transfer
  model for carbonation of concrete}},
\newblock \bibinfo{journal}{Cement and Concrete Research} \bibinfo{volume}{34}
  (\bibinfo{year}{2004}) \bibinfo{pages}{1859--1872}.
%Type = Article
\bibitem[{{Omikrine Metalssi} et~al.(2020){Omikrine Metalssi},
  A{\"{i}}t-Mokhtar, and Turcry}]{OmikrineMetalssi2020}
\bibinfo{author}{O.~{Omikrine Metalssi}},
  \bibinfo{author}{A.~A{\"{i}}t-Mokhtar}, \bibinfo{author}{P.~Turcry},
\newblock \bibinfo{title}{{A proposed modelling of coupling
  carbonation-porosity-moisture transfer in concrete based on mass balance
  equilibrium}},
\newblock \bibinfo{journal}{Construction and Building Materials}
  \bibinfo{volume}{230} (\bibinfo{year}{2020}) \bibinfo{pages}{116997}.
%Type = Article
\bibitem[{young Hwang et~al.(2020)young Hwang, Kwak, and Shim}]{Hwang2020}
\bibinfo{author}{J.~young Hwang}, \bibinfo{author}{H.~G. Kwak},
  \bibinfo{author}{M.~Shim},
\newblock \bibinfo{title}{{Numerical approach for concrete carbonation
  considering moisture diffusion}},
\newblock \bibinfo{journal}{Materials and Structures} \bibinfo{volume}{53}
  (\bibinfo{year}{2020}) \bibinfo{pages}{1--10}.
%Type = Article
\bibitem[{Seigneur et~al.(2020)Seigneur, Kangni-Foli, Lagneau, Dauz{\`{e}}res,
  Poyet, Bescop, L'H{\^{o}}pital, and d'Espinose~de Lacaillerie}]{Seigneur2020}
\bibinfo{author}{N.~Seigneur}, \bibinfo{author}{E.~Kangni-Foli},
  \bibinfo{author}{V.~Lagneau}, \bibinfo{author}{A.~Dauz{\`{e}}res},
  \bibinfo{author}{S.~Poyet}, \bibinfo{author}{P.~L. Bescop},
  \bibinfo{author}{E.~L'H{\^{o}}pital}, \bibinfo{author}{J.~B. d'Espinose~de
  Lacaillerie},
\newblock \bibinfo{title}{{Predicting the atmospheric carbonation of
  cementitious materials using fully coupled two-phase reactive transport
  modelling}},
\newblock \bibinfo{journal}{Cement and Concrete Research} \bibinfo{volume}{130}
  (\bibinfo{year}{2020}) \bibinfo{pages}{105966}.
%Type = Article
\bibitem[{Seigneur et~al.(2022)Seigneur, {De Windt}, Poyet, Soci{\'{e}}, and
  Dauz{\`{e}}res}]{Seigneur2022}
\bibinfo{author}{N.~Seigneur}, \bibinfo{author}{L.~{De Windt}},
  \bibinfo{author}{S.~Poyet}, \bibinfo{author}{A.~Soci{\'{e}}},
  \bibinfo{author}{A.~Dauz{\`{e}}res},
\newblock \bibinfo{title}{{Modelling of the evolving contributions of gas
  transport, cracks and chemical kinetics during atmospheric carbonation of
  hydrated C3S and C-S-H pastes}},
\newblock \bibinfo{journal}{Cement and Concrete Research} \bibinfo{volume}{160}
  (\bibinfo{year}{2022}).
%Type = Article
\bibitem[{Bretti et~al.(2022)Bretti, Ceseri, Natalini, Ciacchella, Santarelli,
  and Tiracorrendo}]{Bretti2022}
\bibinfo{author}{G.~Bretti}, \bibinfo{author}{M.~Ceseri},
  \bibinfo{author}{R.~Natalini}, \bibinfo{author}{M.~C. Ciacchella},
  \bibinfo{author}{M.~L. Santarelli}, \bibinfo{author}{G.~Tiracorrendo},
\newblock \bibinfo{title}{{A forecasting model for the porosity variation
  during the carbonation process}},
\newblock \bibinfo{journal}{GEM - International Journal on Geomathematics}
  \bibinfo{volume}{13} (\bibinfo{year}{2022}) \bibinfo{pages}{1--24}.
%Type = Article
\bibitem[{Nguyen et~al.(2015)Nguyen, Bary, and {De Larrard}}]{Nguyen2015}
\bibinfo{author}{T.~T. Nguyen}, \bibinfo{author}{B.~Bary},
  \bibinfo{author}{T.~{De Larrard}},
\newblock \bibinfo{title}{{Coupled carbonation-rust formation-damage modeling
  and simulation of steel corrosion in 3D mesoscale reinforced concrete}},
\newblock \bibinfo{journal}{Cement and Concrete Research} \bibinfo{volume}{74}
  (\bibinfo{year}{2015}) \bibinfo{pages}{95--107}.
%Type = Article
\bibitem[{Forsdyke and Lees(2023)}]{Forsdyke2023}
\bibinfo{author}{J.~C. Forsdyke}, \bibinfo{author}{J.~M. Lees},
\newblock \bibinfo{title}{{Model fitting to concrete carbonation data with
  non-zero initial carbonation depth}},
\newblock \bibinfo{journal}{Materials and Structures} \bibinfo{volume}{56}
  (\bibinfo{year}{2023}) \bibinfo{pages}{1--11}.
%Type = Article
\bibitem[{Phung et~al.(2016)Phung, Maes, Jacques, {De Schutter}, Ye, and
  Perko}]{Phung2016}
\bibinfo{author}{Q.~T. Phung}, \bibinfo{author}{N.~Maes},
  \bibinfo{author}{D.~Jacques}, \bibinfo{author}{G.~{De Schutter}},
  \bibinfo{author}{G.~Ye}, \bibinfo{author}{J.~Perko},
\newblock \bibinfo{title}{{Modelling the carbonation of cement pastes under a
  CO2 pressure gradient considering both diffusive and convective transport}},
\newblock \bibinfo{journal}{Construction and Building Materials}
  \bibinfo{volume}{114} (\bibinfo{year}{2016}) \bibinfo{pages}{333--351}.
%Type = Article
\bibitem[{Kari et~al.(2014)Kari, Puttonen, and Skantz}]{Kari2014}
\bibinfo{author}{O.~P. Kari}, \bibinfo{author}{J.~Puttonen},
  \bibinfo{author}{E.~Skantz},
\newblock \bibinfo{title}{{Reactive transport modelling of long-term
  carbonation}},
\newblock \bibinfo{journal}{Cement and Concrete Composites}
  \bibinfo{volume}{52} (\bibinfo{year}{2014}) \bibinfo{pages}{42--53}.
%Type = Article
\bibitem[{Zhu et~al.(2016)Zhu, Zi, Cao, and Cheng}]{Zhu2016a}
\bibinfo{author}{X.~Zhu}, \bibinfo{author}{G.~Zi}, \bibinfo{author}{Z.~Cao},
  \bibinfo{author}{X.~Cheng},
\newblock \bibinfo{title}{{Combined effect of carbonation and chloride ingress
  in concrete}},
\newblock \bibinfo{journal}{Construction and Building Materials}
  \bibinfo{volume}{110} (\bibinfo{year}{2016}) \bibinfo{pages}{369--380}.
%Type = Article
\bibitem[{Han~Shen et~al.(2019)Han~Shen, qiang Jiang, Hou, Hu, Yang, and feng
  Liu}]{Shen2019}
\bibinfo{author}{X.~Han~Shen}, \bibinfo{author}{W.~qiang Jiang},
  \bibinfo{author}{D.~Hou}, \bibinfo{author}{Z.~Hu}, \bibinfo{author}{J.~Yang},
  \bibinfo{author}{Q.~feng Liu},
\newblock \bibinfo{title}{{Numerical study of carbonation and its effect on
  chloride binding in concrete}},
\newblock \bibinfo{journal}{Cement and Concrete Composites}
  \bibinfo{volume}{104} (\bibinfo{year}{2019}) \bibinfo{pages}{103402}.
%Type = Article
\bibitem[{Meijers et~al.(2005)Meijers, Bijen, {De Borst}, and
  Fraaij}]{Meijers2005}
\bibinfo{author}{S.~J. Meijers}, \bibinfo{author}{J.~M. Bijen},
  \bibinfo{author}{R.~{De Borst}}, \bibinfo{author}{A.~L. Fraaij},
\newblock \bibinfo{title}{{Computational results of a model for chloride
  ingress in concrete including convection, drying-wetting cycles and
  carbonation}},
\newblock \bibinfo{journal}{Materials and Structures} \bibinfo{volume}{38}
  (\bibinfo{year}{2005}) \bibinfo{pages}{145--154}.
%Type = Article
\bibitem[{Li et~al.(2018)Li, Zhang, Wang, and Zeng}]{Li2018}
\bibinfo{author}{K.~Li}, \bibinfo{author}{Y.~Zhang}, \bibinfo{author}{S.~Wang},
  \bibinfo{author}{J.~Zeng},
\newblock \bibinfo{title}{{Impact of carbonation on the chloride diffusivity in
  concrete: experiment, analysis and application}},
\newblock \bibinfo{journal}{Materials and Structures} \bibinfo{volume}{51}
  (\bibinfo{year}{2018}) \bibinfo{pages}{1--15}.
%Type = Article
\bibitem[{Xie et~al.(2021)Xie, Dangla, and Li}]{Xie2021}
\bibinfo{author}{M.~Xie}, \bibinfo{author}{P.~Dangla}, \bibinfo{author}{K.~Li},
\newblock \bibinfo{title}{{Reactive transport modelling of concrete subject to
  de-icing salts and atmospheric carbonation}},
\newblock \bibinfo{journal}{Materials and Structures} \bibinfo{volume}{54}
  (\bibinfo{year}{2021}) \bibinfo{pages}{1--15}.
%Type = Article
\bibitem[{Freddi and Mingazzi(2022)}]{Freddi2022}
\bibinfo{author}{F.~Freddi}, \bibinfo{author}{L.~Mingazzi},
\newblock \bibinfo{title}{{A predictive phase-field approach for cover cracking
  in corroded concrete elements}},
\newblock \bibinfo{journal}{Theoretical and Applied Fracture Mechanics}
  \bibinfo{volume}{122} (\bibinfo{year}{2022}) \bibinfo{pages}{103657}.
%Type = Article
\bibitem[{Grassl(2009)}]{Grassl2009}
\bibinfo{author}{P.~Grassl},
\newblock \bibinfo{title}{{A lattice approach to model flow in cracked
  concrete}},
\newblock \bibinfo{journal}{Cement and Concrete Composites}
  \bibinfo{volume}{31} (\bibinfo{year}{2009}) \bibinfo{pages}{454--460}.
%Type = Article
\bibitem[{Stefanoni et~al.(2019)Stefanoni, Angst, and Elsener}]{Stefanoni2019}
\bibinfo{author}{M.~Stefanoni}, \bibinfo{author}{U.~M. Angst},
  \bibinfo{author}{B.~Elsener},
\newblock \bibinfo{title}{{Kinetics of electrochemical dissolution of metals in
  porous media}},
\newblock \bibinfo{journal}{Nature Materials} \bibinfo{volume}{18}
  (\bibinfo{year}{2019}) \bibinfo{pages}{942--947}.
%Type = Article
\bibitem[{Mainguy et~al.(2001)Mainguy, Coussy, and
  Baroghel-Bouny}]{mainguy2001role}
\bibinfo{author}{M.~Mainguy}, \bibinfo{author}{O.~Coussy},
  \bibinfo{author}{V.~Baroghel-Bouny},
\newblock \bibinfo{title}{{Role of air pressure in drying of weakly permeable
  materials}},
\newblock \bibinfo{journal}{Journal of engineering mechanics}
  \bibinfo{volume}{127} (\bibinfo{year}{2001}) \bibinfo{pages}{582--592}.
%Type = Article
\bibitem[{{Van Genuchten}(1980)}]{van1980closed}
\bibinfo{author}{M.~T. {Van Genuchten}},
\newblock \bibinfo{title}{{A closed-form equation for predicting the hydraulic
  conductivity of unsaturated soils}},
\newblock \bibinfo{journal}{Soil science society of America journal}
  \bibinfo{volume}{44} (\bibinfo{year}{1980}) \bibinfo{pages}{892--898}.
%Type = Article
\bibitem[{Zhang et~al.(2015)Zhang, Thiery, and Baroghel-Bouny}]{Zhang2015a}
\bibinfo{author}{Z.~Zhang}, \bibinfo{author}{M.~Thiery},
  \bibinfo{author}{V.~Baroghel-Bouny},
\newblock \bibinfo{title}{{Numerical modelling of moisture transfers with
  hysteresis within cementitious materials: Verification and investigation of
  the effects of repeated wetting-drying boundary conditions}},
\newblock \bibinfo{journal}{Cement and Concrete Research} \bibinfo{volume}{68}
  (\bibinfo{year}{2015}) \bibinfo{pages}{10--23}.
%Type = Article
\bibitem[{Mualem(1976)}]{mualem1976new}
\bibinfo{author}{Y.~Mualem},
\newblock \bibinfo{title}{{A new model for predicting the hydraulic
  conductivity of unsaturated porous media}},
\newblock \bibinfo{journal}{Water resources research} \bibinfo{volume}{12}
  (\bibinfo{year}{1976}) \bibinfo{pages}{513--522}.
%Type = Article
\bibitem[{Miehe et~al.(2015)Miehe, Mauthe, and Teichtmeister}]{Miehe2015b}
\bibinfo{author}{C.~Miehe}, \bibinfo{author}{S.~Mauthe},
  \bibinfo{author}{S.~Teichtmeister},
\newblock \bibinfo{title}{{Minimization principles for the coupled problem of
  Darcy-Biot-type fluid transport in porous media linked to phase field
  modeling of fracture}},
\newblock \bibinfo{journal}{Journal of the Mechanics and Physics of Solids}
  \bibinfo{volume}{82} (\bibinfo{year}{2015}) \bibinfo{pages}{186--217}.
%Type = Article
\bibitem[{Wilson and Landis(2016)}]{Wilson2016}
\bibinfo{author}{Z.~A. Wilson}, \bibinfo{author}{C.~M. Landis},
\newblock \bibinfo{title}{{Phase-field modeling of hydraulic fracture}},
\newblock \bibinfo{journal}{Journal of the Mechanics and Physics of Solids}
  \bibinfo{volume}{96} (\bibinfo{year}{2016}) \bibinfo{pages}{264--290}.
%Type = Article
\bibitem[{Heider and Sun(2020)}]{Heider2020}
\bibinfo{author}{Y.~Heider}, \bibinfo{author}{W.~C. Sun},
\newblock \bibinfo{title}{{A phase field framework for capillary-induced
  fracture in unsaturated porous media: Drying-induced vs. hydraulic
  cracking}},
\newblock \bibinfo{journal}{Computer Methods in Applied Mechanics and
  Engineering} \bibinfo{volume}{359} (\bibinfo{year}{2020})
  \bibinfo{pages}{112647}.
%Type = Article
\bibitem[{Zhang et~al.(2022)Zhang, Trtik, Ren, Schmid, Dreimol, and
  Angst}]{Zhang2022}
\bibinfo{author}{Z.~Zhang}, \bibinfo{author}{P.~Trtik},
  \bibinfo{author}{F.~Ren}, \bibinfo{author}{T.~Schmid}, \bibinfo{author}{C.~H.
  Dreimol}, \bibinfo{author}{U.~Angst},
\newblock \bibinfo{title}{{Dynamic effect of water penetration on steel
  corrosion in carbonated mortar: A neutron imaging, electrochemical, and
  modeling study}},
\newblock \bibinfo{journal}{Cement} \bibinfo{volume}{9} (\bibinfo{year}{2022})
  \bibinfo{pages}{100043}.
%Type = Article
\bibitem[{Wu and Chen(2021)}]{Wu2021}
\bibinfo{author}{J.-Y. Wu}, \bibinfo{author}{W.~X. Chen},
\newblock \bibinfo{title}{{Phase-field modeling of electromechanical fracture
  in piezoelectric solids: Analytical results and numerical simulations}},
\newblock \bibinfo{journal}{Computer Methods in Applied Mechanics and
  Engineering} \bibinfo{volume}{387} (\bibinfo{year}{2021})
  \bibinfo{pages}{114125}.
%Type = Article
\bibitem[{Kristensen et~al.(2021)Kristensen, Niordson, and
  Mart{\'{i}}nez-Pa{\~{n}}eda}]{Kristensen2021}
\bibinfo{author}{P.~K. Kristensen}, \bibinfo{author}{C.~F. Niordson},
  \bibinfo{author}{E.~Mart{\'{i}}nez-Pa{\~{n}}eda},
\newblock \bibinfo{title}{{An assessment of phase field fracture: Crack
  initiation and growth}},
\newblock \bibinfo{journal}{Philosophical Transactions of the Royal Society A:
  Mathematical, Physical and Engineering Sciences} \bibinfo{volume}{379}
  (\bibinfo{year}{2021}) \bibinfo{pages}{20210021}.
%Type = Article
\bibitem[{Miehe et~al.(2010)Miehe, Welschinger, and Hofacker}]{Miehe2010}
\bibinfo{author}{C.~Miehe}, \bibinfo{author}{F.~Welschinger},
  \bibinfo{author}{M.~Hofacker},
\newblock \bibinfo{title}{{Thermodynamically consistent phase-field models of
  fracture: Variational principles and multi-field FE implementations}},
\newblock \bibinfo{journal}{International Journal for Numerical Methods in
  Engineering} \bibinfo{volume}{83} (\bibinfo{year}{2010})
  \bibinfo{pages}{1273--1311}.
%Type = Article
\bibitem[{Amor et~al.(2009)Amor, Marigo, and Maurini}]{Amor2009}
\bibinfo{author}{H.~Amor}, \bibinfo{author}{J.~J. Marigo},
  \bibinfo{author}{C.~Maurini},
\newblock \bibinfo{title}{{Regularized formulation of the variational brittle
  fracture with unilateral contact: Numerical experiments}},
\newblock \bibinfo{journal}{Journal of the Mechanics and Physics of Solids}
  \bibinfo{volume}{57} (\bibinfo{year}{2009}) \bibinfo{pages}{1209--1229}.
%Type = Article
\bibitem[{Bourdin et~al.(2000)Bourdin, Francfort, and Marigo}]{Bourdin2000}
\bibinfo{author}{B.~Bourdin}, \bibinfo{author}{G.~A. Francfort},
  \bibinfo{author}{J.~J. Marigo},
\newblock \bibinfo{title}{{Numerical experiments in revisited brittle
  fracture}},
\newblock \bibinfo{journal}{Journal of the Mechanics and Physics of Solids}
  \bibinfo{volume}{48} (\bibinfo{year}{2000}) \bibinfo{pages}{797--826}.
%Type = Article
\bibitem[{Korec et~al.(2023)Korec, Jir{\'a}sek, Wong, and
  Mart{\'\i}nez-Pa{\~n}eda}]{Korec2023}
\bibinfo{author}{E.~Korec}, \bibinfo{author}{M.~Jir{\'a}sek},
  \bibinfo{author}{H.~S. Wong}, \bibinfo{author}{E.~Mart{\'\i}nez-Pa{\~n}eda},
\newblock \bibinfo{title}{A phase-field chemo-mechanical model for
  corrosion-induced cracking in reinforced concrete},
\newblock \bibinfo{journal}{Construction and Building Materials}
  \bibinfo{volume}{393} (\bibinfo{year}{2023}) \bibinfo{pages}{131964}.
%Type = Article
\bibitem[{Carrara et~al.(2020)Carrara, Ambati, Alessi, and {De
  Lorenzis}}]{Carrara2020}
\bibinfo{author}{P.~Carrara}, \bibinfo{author}{M.~Ambati},
  \bibinfo{author}{R.~Alessi}, \bibinfo{author}{L.~{De Lorenzis}},
\newblock \bibinfo{title}{{A framework to model the fatigue behavior of brittle
  materials based on a variational phase-field approach}},
\newblock \bibinfo{journal}{Computer Methods in Applied Mechanics and
  Engineering} \bibinfo{volume}{361} (\bibinfo{year}{2020})
  \bibinfo{pages}{112731}.
%Type = Article
\bibitem[{Navidtehrani et~al.(2022)Navidtehrani, Beteg{\'{o}}n, and
  Mart{\'{i}}nez-Pa{\~{n}}eda}]{Navidtehrani2022}
\bibinfo{author}{Y.~Navidtehrani}, \bibinfo{author}{C.~Beteg{\'{o}}n},
  \bibinfo{author}{E.~Mart{\'{i}}nez-Pa{\~{n}}eda},
\newblock \bibinfo{title}{{A general framework for decomposing the phase field
  fracture driving force, particularised to a Drucker–Prager failure
  surface}},
\newblock \bibinfo{journal}{Theoretical and Applied Fracture Mechanics}
  \bibinfo{volume}{121} (\bibinfo{year}{2022}) \bibinfo{pages}{103555}.
%Type = Article
\bibitem[{jie Huang et~al.(2022)jie Huang, jun Yang, Zhang, and
  Natarajan}]{Huang2022}
\bibinfo{author}{Y.~jie Huang}, \bibinfo{author}{Z.~jun Yang},
  \bibinfo{author}{H.~Zhang}, \bibinfo{author}{S.~Natarajan},
\newblock \bibinfo{title}{{A phase-field cohesive zone model integrated with
  cell-based smoothed finite element method for quasi-brittle fracture
  simulations of concrete at mesoscale}},
\newblock \bibinfo{journal}{Computer Methods in Applied Mechanics and
  Engineering} \bibinfo{volume}{396} (\bibinfo{year}{2022})
  \bibinfo{pages}{115074}.
%Type = Article
\bibitem[{Wu(2017)}]{Wu2017}
\bibinfo{author}{J.-Y. Wu},
\newblock \bibinfo{title}{{A unified phase-field theory for the mechanics of
  damage and quasi-brittle failure}},
\newblock \bibinfo{journal}{Journal of the Mechanics and Physics of Solids}
  \bibinfo{volume}{103} (\bibinfo{year}{2017}) \bibinfo{pages}{72--99}.
%Type = Article
\bibitem[{Wu(2018)}]{Wu2018a}
\bibinfo{author}{J.-Y. Wu},
\newblock \bibinfo{title}{{Robust numerical implementation of non-standard
  phase-field damage models for failure in solids}},
\newblock \bibinfo{journal}{Computer Methods in Applied Mechanics and
  Engineering} \bibinfo{volume}{340} (\bibinfo{year}{2018})
  \bibinfo{pages}{767--797}.
%Type = Article
\bibitem[{Pundir et~al.(2023)Pundir, Kammer, and Angst}]{Pundir2023}
\bibinfo{author}{M.~Pundir}, \bibinfo{author}{D.~S. Kammer},
  \bibinfo{author}{U.~Angst},
\newblock \bibinfo{title}{{Journal of the Mechanics and Physics of Solids An
  FFT-based framework for predicting corrosion-driven damage in fractal porous
  media}},
\newblock \bibinfo{journal}{Journal of the Mechanics and Physics of Solids}
  \bibinfo{volume}{179} (\bibinfo{year}{2023}) \bibinfo{pages}{105388}.
%Type = Article
\bibitem[{Wu and Chen(2022)}]{Wu2022}
\bibinfo{author}{J.-Y. Wu}, \bibinfo{author}{W.~X. Chen},
\newblock \bibinfo{title}{{On the phase-field modeling of fully coupled
  chemo-mechanical deterioration and fracture in calcium leached cementitious
  solids}},
\newblock \bibinfo{journal}{International Journal of Solids and Structures}
  \bibinfo{volume}{238} (\bibinfo{year}{2022}) \bibinfo{pages}{111380}.
%Type = Article
\bibitem[{Korec et~al.(2024)Korec, Jir{\'a}sek, Wong, and
  Mart{\'\i}nez-Pa{\~n}eda}]{korec2024phase}
\bibinfo{author}{E.~Korec}, \bibinfo{author}{M.~Jir{\'a}sek},
  \bibinfo{author}{H.~S. Wong}, \bibinfo{author}{E.~Mart{\'\i}nez-Pa{\~n}eda},
\newblock \bibinfo{title}{Phase-field chemo-mechanical modelling of
  corrosion-induced cracking in reinforced concrete subjected to non-uniform
  chloride-induced corrosion},
\newblock \bibinfo{journal}{Theoretical and Applied Fracture Mechanics}
  \bibinfo{volume}{129} (\bibinfo{year}{2024}) \bibinfo{pages}{104233}.
%Type = Article
\bibitem[{Angst et~al.(2017)Angst, Geiker, Michel, Gehlen, Wong, Isgor,
  Elsener, Hansson, Fran{\c{c}}ois, Hornbostel, Polder, Alonso, Sanchez,
  Correia, Criado, Sag{\"{u}}{\'{e}}s, and Buenfeld}]{Angst2017}
\bibinfo{author}{U.~M. Angst}, \bibinfo{author}{M.~R. Geiker},
  \bibinfo{author}{A.~Michel}, \bibinfo{author}{C.~Gehlen},
  \bibinfo{author}{H.~Wong}, \bibinfo{author}{O.~B. Isgor},
  \bibinfo{author}{B.~Elsener}, \bibinfo{author}{C.~M. Hansson},
  \bibinfo{author}{R.~Fran{\c{c}}ois}, \bibinfo{author}{K.~Hornbostel},
  \bibinfo{author}{R.~Polder}, \bibinfo{author}{M.~C. Alonso},
  \bibinfo{author}{M.~Sanchez}, \bibinfo{author}{M.~J. Correia},
  \bibinfo{author}{M.~Criado}, \bibinfo{author}{A.~Sag{\"{u}}{\'{e}}s},
  \bibinfo{author}{N.~Buenfeld},
\newblock \bibinfo{title}{{The steel–concrete interface}},
\newblock \bibinfo{journal}{Materials and Structures} \bibinfo{volume}{50}
  (\bibinfo{year}{2017}).
%Type = Article
\bibitem[{Wong et~al.(2022)Wong, Angst, Geiker, Isgor, Elsener, Michel, Alonso,
  Correia, Pacheco, Gulikers, Zhao, Criado, Raupach, S{\o}rensen,
  Fran{\c{c}}ois, Mundra, Rasol, and Polder}]{Wong2022}
\bibinfo{author}{H.~S. Wong}, \bibinfo{author}{U.~M. Angst},
  \bibinfo{author}{M.~R. Geiker}, \bibinfo{author}{O.~B. Isgor},
  \bibinfo{author}{B.~Elsener}, \bibinfo{author}{A.~Michel},
  \bibinfo{author}{M.~C. Alonso}, \bibinfo{author}{M.~J. Correia},
  \bibinfo{author}{J.~Pacheco}, \bibinfo{author}{J.~Gulikers},
  \bibinfo{author}{Y.~Zhao}, \bibinfo{author}{M.~Criado},
  \bibinfo{author}{M.~Raupach}, \bibinfo{author}{H.~S{\o}rensen},
  \bibinfo{author}{R.~Fran{\c{c}}ois}, \bibinfo{author}{S.~Mundra},
  \bibinfo{author}{M.~Rasol}, \bibinfo{author}{R.~Polder},
\newblock \bibinfo{title}{{Methods for characterising the steel–concrete
  interface to enhance understanding of reinforcement corrosion: a critical
  review by RILEM TC 262-SCI}},
\newblock \bibinfo{journal}{Materials and Structures} \bibinfo{volume}{55}
  (\bibinfo{year}{2022}).
%Type = Article
\bibitem[{Arndt et~al.(2021)Arndt, Bangerth, Davydov, Heister, Heltai,
  Kronbichler, Maier, Pelteret, Turcksin, and Wells}]{Arndt2021}
\bibinfo{author}{D.~Arndt}, \bibinfo{author}{W.~Bangerth},
  \bibinfo{author}{D.~Davydov}, \bibinfo{author}{T.~Heister},
  \bibinfo{author}{L.~Heltai}, \bibinfo{author}{M.~Kronbichler},
  \bibinfo{author}{M.~Maier}, \bibinfo{author}{J.~P. Pelteret},
  \bibinfo{author}{B.~Turcksin}, \bibinfo{author}{D.~Wells},
\newblock \bibinfo{title}{{The DEAL.II finite element library: Design,
  features, and insights}},
\newblock \bibinfo{journal}{Computers and Mathematics with Applications}
  \bibinfo{volume}{81} (\bibinfo{year}{2021}) \bibinfo{pages}{407--422}.
%Type = Article
\bibitem[{Bangerth et~al.(2007)Bangerth, Hartmann, and Kanschat}]{Bangerth2007}
\bibinfo{author}{W.~Bangerth}, \bibinfo{author}{R.~Hartmann},
  \bibinfo{author}{G.~Kanschat},
\newblock \bibinfo{title}{{Deal.II - A general-purpose object-oriented finite
  element library}},
\newblock \bibinfo{journal}{ACM Transactions on Mathematical Software}
  \bibinfo{volume}{33} (\bibinfo{year}{2007}) \bibinfo{pages}{1--27}.
%Type = Article
\bibitem[{Baroghel-Bouny et~al.(1999)Baroghel-Bouny, Mainguy, Lassabatere, and
  Coussy}]{Baroghel-Bouny1999}
\bibinfo{author}{V.~Baroghel-Bouny}, \bibinfo{author}{M.~Mainguy},
  \bibinfo{author}{T.~Lassabatere}, \bibinfo{author}{O.~Coussy},
\newblock \bibinfo{title}{{Characterization and identification of equilibrium
  and transfer moisture properties for ordinary and high-performance
  cementitious materials}},
\newblock \bibinfo{journal}{Cement and Concrete Research} \bibinfo{volume}{29}
  (\bibinfo{year}{1999}) \bibinfo{pages}{1225--1238}.
%Type = Article
\bibitem[{Michel and Pease(2018)}]{Michel2018}
\bibinfo{author}{A.~Michel}, \bibinfo{author}{B.~J. Pease},
\newblock \bibinfo{title}{{Moisture ingress in cracked cementitious
  materials}},
\newblock \bibinfo{journal}{Cement and Concrete Research} \bibinfo{volume}{113}
  (\bibinfo{year}{2018}) \bibinfo{pages}{154--168}.
%Type = Article
\bibitem[{Liu et~al.(2020)Liu, Yu, and Chen}]{Liu2020}
\bibinfo{author}{P.~Liu}, \bibinfo{author}{Z.~Yu}, \bibinfo{author}{Y.~Chen},
\newblock \bibinfo{title}{{Carbonation depth model and carbonated acceleration
  rate of concrete under different environment}},
\newblock \bibinfo{journal}{Cement and Concrete Composites}
  \bibinfo{volume}{114} (\bibinfo{year}{2020}) \bibinfo{pages}{103736}.
%Type = Article
\bibitem[{Zhang et~al.(2016)Zhang, Thiery, and Baroghel-Bouny}]{Zhang2016a}
\bibinfo{author}{Z.~Zhang}, \bibinfo{author}{M.~Thiery},
  \bibinfo{author}{V.~Baroghel-Bouny},
\newblock \bibinfo{title}{{Investigation of moisture transport properties of
  cementitious materials}},
\newblock \bibinfo{journal}{Cement and Concrete Research} \bibinfo{volume}{89}
  (\bibinfo{year}{2016}) \bibinfo{pages}{257--268}.
%Type = Article
\bibitem[{Powers and Brownyard(1946)}]{Powers1946}
\bibinfo{author}{T.~C. Powers}, \bibinfo{author}{T.~L. Brownyard},
\newblock \bibinfo{title}{{Studies of the physical properties of hardened
  portland cement paste}},
\newblock \bibinfo{journal}{American Concrete Institute, ACI Special
  Publication} \bibinfo{volume}{SP-249} (\bibinfo{year}{1946})
  \bibinfo{pages}{265--617}.
%Type = Article
\bibitem[{Andrade et~al.(1999)Andrade, Sarr{\'{i}}a, and Alonso}]{Andrade1999}
\bibinfo{author}{C.~Andrade}, \bibinfo{author}{J.~Sarr{\'{i}}a},
  \bibinfo{author}{C.~Alonso},
\newblock \bibinfo{title}{{Relative humidity in the interior of concrete
  exposed to natural and artificial weathering}},
\newblock \bibinfo{journal}{Cement and Concrete Research} \bibinfo{volume}{29}
  (\bibinfo{year}{1999}) \bibinfo{pages}{1249--1259}.
%Type = Article
\bibitem[{{Angulo Ramirez} et~al.(2023){Angulo Ramirez}, Meira, Quattrone, and
  John}]{AnguloRamirez2023}
\bibinfo{author}{D.~E. {Angulo Ramirez}}, \bibinfo{author}{G.~R. Meira},
  \bibinfo{author}{M.~Quattrone}, \bibinfo{author}{V.~M. John},
\newblock \bibinfo{title}{{A review on reinforcement corrosion propagation in
  carbonated concrete – Influence of material and environmental
  characteristics}},
\newblock \bibinfo{journal}{Cement and Concrete Composites}
  \bibinfo{volume}{140} (\bibinfo{year}{2023}) \bibinfo{pages}{105085}.
%Type = Book
\bibitem[{Andrade(2023)}]{Andrade2023a}
\bibinfo{author}{C.~Andrade}, \bibinfo{title}{{Role of Oxygen and Humidity in
  the Reinforcement Corrosion}}, \bibinfo{publisher}{Springer International
  Publishing}, \bibinfo{year}{2023}.
%Type = Article
\bibitem[{Hageman et~al.(2023)Hageman, Andrade, and
  Mart{\'\i}nez-Pa{\~n}eda}]{hageman2023corrosion}
\bibinfo{author}{T.~Hageman}, \bibinfo{author}{C.~Andrade},
  \bibinfo{author}{E.~Mart{\'\i}nez-Pa{\~n}eda},
\newblock \bibinfo{title}{Corrosion rates under charge-conservation
  conditions},
\newblock \bibinfo{journal}{Electrochimica Acta} \bibinfo{volume}{461}
  (\bibinfo{year}{2023}) \bibinfo{pages}{142624}.

\end{thebibliography}

\end{document}